\documentclass[pdflatex,iicol,sn-mathphys-num]{sn-jnl}
\usepackage{graphicx}
\graphicspath{{./}}

\usepackage{latexsym}
\usepackage{amsmath}
\usepackage{amssymb}

\usepackage[utf8]{inputenc}
\hypersetup{
    colorlinks=true,
    linkcolor=blue,
    citecolor=blue,
    pdftitle={Clock-noise propagation and calibration for phase-locking configurations in space-based gravitational-wave detectors},
    pdfauthor={Pan-Pan Wang and Cheng-Gang Shao},
}
\usepackage{color}
\usepackage[T1]{fontenc}

\usepackage{microtype}
\usepackage{supertabular}
\usepackage{setspace}
\usepackage{multirow}
\usepackage{makecell}
\usepackage{mathrsfs}

\usepackage[center]{subfigure}
\makeatletter
\newenvironment{inlinefigure}{%
  \par\addvspace{8pt}\noindent
  \begin{minipage}{\columnwidth}\def\@captype{figure}%
}{%
  \end{minipage}\par\addvspace{8pt}%
}
\makeatother
\usepackage{multicol}
\raggedcolumns
 \newcommand{\bq}{\begin{equation}}
 \newcommand{\eq}{\end{equation}}
 \newcommand{\bqn}{\begin{eqnarray}}
 \newcommand{\eqn}{\end{eqnarray}}

\makeatletter

\newcommand{\Rmnum}[1]{\expandafter\@slowromancap\romannumeral #1@}

\makeatother

\allowdisplaybreaks[1]
\makeatletter
\g@addto@macro\normalsize{%
  \abovedisplayskip 8pt plus 2pt minus 1pt%
  \belowdisplayskip 8pt plus 2pt minus 1pt%
}
\makeatother
\begin{document}
\onecolumn
\raggedbottom

\title{Clock-noise propagation and calibration for phase-locking configurations in space-based gravitational-wave detectors}

\author[1]{\fnm{Pan-Pan} \sur{Wang}}
\author[2]{\fnm{Cheng-Gang} \sur{Shao}}
\affil[1]{\orgdiv{College of Physics}, \orgname{Chongqing University}, \orgaddress{\city{Chongqing}, \postcode{401331}, \country{China}}}
\affil[2]{\orgdiv{School of Physics and Optoelectronic Engineering}, \orgname{Yangtze University}, \orgaddress{\city{Jingzhou}, \postcode{434023}, \country{China}}}

\abstract{In space-based gravitational-wave detectors, laser phase locking transfers the phase of weak received light to a local laser and maintains the interspacecraft heterodyne frequencies within the phasemeter bandwidth. Fluctuations of the onboard ultrastable oscillators introduce clock noise into these heterodyne phase measurements, requiring additional calibration in time-delay interferometry (TDI). Existing formulations based on six independently formed one-way measurements do not explicitly describe the common-reference clock structure of a master--slave phase-locking configuration. In this work, we formulate clock-noise propagation and calibration directly for this configuration. Starting from carrier and sideband readouts, we derive the clock-noise couplings and construct an ordered sideband-calibration rule. The rule retains noncommuting delays and applies to two-branch TDI combinations satisfying propagation and clock-coefficient closure. For mutually independent onboard clocks with identical fractional-frequency noise spectra, we compare matched realizations with and without phase locking. In the adopted frozen equal-arm model, 45 second-generation combinations with up to 16 links considered here have smaller clock-noise residuals in the phase-locking realization throughout 0.1--10~mHz, before final clock calibration. Numerical simulations of two representative 16-link combinations reproduce the analytical spectra and verify the predicted reduction and calibration. These results show that the phase-locking realizations reduce the clock-noise contribution before final calibration, while the cancellation of master-laser phase noise still relies on TDI.}



\maketitle
\renewcommand{\dbltopfraction}{0.8}\renewcommand{\textfraction}{0.05}\setcounter{dbltopnumber}{1}

\begin{multicols}{2}
\section{Introduction}\label{section1}
The Laser Interferometer Space Antenna (LISA)~\cite{gw-lisa1,gw-lisa2},
Taiji~\cite{gw-Taiji}, and TianQin~\cite{gw-tianqin} are designed to observe
gravitational waves (GWs) in the millihertz band, which is inaccessible to
terrestrial interferometers.  Their long baselines will enable observations
of massive black-hole binaries, extreme-mass-ratio inspirals, and compact
Galactic binaries.  The long-lived signals from these systems can probe the
assembly of massive black holes and strong-field gravity, while supporting
advance source localization and coordinated multimessenger observations.
Realizing this scientific potential places stringent demands on long-baseline
interferometric measurement and instrumental-noise suppression.

Laser phase noise is approximately seven orders of magnitude larger than the
target GW signals.  The unequal and time-dependent arm lengths prevent its
direct common-mode cancellation.  Time-delay interferometry (TDI)
~\cite{tdi-01,tdi-02,tdi-03} combines appropriately delayed measurements to
synthesize nearly equal optical paths and suppress this noise.
After more than 20 years of development, the performance of TDI in mitigating laser phase noise has been well validated both theoretically~\cite{frame-01-2000,tdi-d55-2001,res-semi--01-2002,tdi-Algebraic-2002,tdi-d22,tdi-laser-01,tdi-d99,tdi-d88, tdi-laser-06,tdi-laser-LISACode,tdi-2010-Dhurandhar, tdi-otto-2015, tdi-filter-s4, algebra-tdi-Wu,algebra-tdi-Qian} and experimentally~\cite{TDIexper-deVine-2010, TDIexper-Vinckier-2020}.

After TDI suppresses laser phase noise, clock noise caused by ultrastable oscillator (USO) fluctuations still exceeds the GW signal by two to three orders of magnitude~\cite{tdi-clock-2001, tdi-clock-2002,tdi-clock-2012,tdi-clock-2015,tdi-clock-2018,tdi-clock-2021,tdi-clock-pan}.
Therefore, a clock sideband comparison TDI algorithm for suppressing clock noise has been developed.
The clock sideband comparison TDI technique utilizes electro-optic modulators to generate sidebands, which carry the remote clock noise information to the local spacecraft.
By differencing the sideband data stream with the carrier data stream, the comparison information between the remote clock noise and the local clock noise is extracted.
The obtained independent clock noise observables are then combined with the appropriate time delay to calibrate the clock noise.
This is referred to as the clock noise reduction scheme, the essence of which is to find the relationship between the additional independent clock noise observables and the residual clock noise after TDI has suppressed the laser phase noise.

Clock-noise calibration was introduced for a static Michelson interferometer
and subsequently implemented in the time domain for first-generation
Michelson and Sagnac observables~\cite{tdi-clock-1996,tdi-clock-2001}.
Later work extended the construction to common first-generation combinations
and to second-generation Michelson, Sagnac, and more general observables
~\cite{tdi-clock-2002,tdi-clock-2018,tdi-clock-2021,tdi-clock-pan}.
Geometric formulations have since exposed reusable clock-calibration patterns
across broader classes of combinations~\cite{clock-2023-Yang}.

Laser phase-locking determines the optical-frequency distribution and the
set of independent phase measurements available to a space-based detector
~\cite{phase-locking-tdi-2024,unitmodelsim-2023}.  Existing clock-noise
calibration methods instead begin with six independently formed one-way
measurements and do not explicitly track the common-reference clock transfer structure imposed by a
master--slave topology.  This difference motivates a formulation built
directly from the phase-locking measurement architecture.

In this work, we formulate clock-noise propagation and calibration directly
for a phase-locking configuration.  We express the carrier and sideband
measurements, together with their laser and clock sectors, in a common-reference
basis.  Signed composite delay operators then provide an ordered calibration
template for time-dependent, noncommuting delays whenever the propagation and
clock-coefficient closure conditions are satisfied.  The construction applies
to every admissible two-branch combination in this class.  Phase-locking
reorganizes the six laser phase processes as delayed copies of a single master
process but does not attenuate the master-laser fluctuation, whose cancellation
still relies on TDI.  We compare the clock-noise transfer for the 45 distinct
second-generation geometric-TDI combinations with up to 16 links considered
here~\cite{geome-tdi-2023} and show that the phase-locking realizations yield
smaller clock-noise residuals throughout $0.1$--$10~\mathrm{mHz}$.
Numerical simulations of $[X]^{16}_{1}$ and $[PE]^{16}_{1}$ provide
representative numerical tests of the analytical transfer functions and the
calibration algorithm.

This paper is organized as follows.
Section~\ref{section2} defines the interferometric measurements and the
intermediary variables used without phase-locking.  Section~\ref{section3}
derives the reduced measurements in the phase-locking configuration.
Section~\ref{section4} develops the signed composite-delay representation.
Section~\ref{section5} constructs the clock-calibration algorithm.
Section~\ref{section6} presents the spectral analysis and numerical
simulations.  Section~\ref{section7} summarizes the results.

\section{Interferometric measurements}\label{section2}

This section establishes the measurement model used throughout the analysis.
We first define the raw carrier, sideband, test-mass, and reference readouts,
and then summarize the intermediary measurements and clock comparisons formed
before phase-locking is imposed.

\subsection{Raw readouts and notation}\label{section2.1}

\begin{inlinefigure}
\centering
\includegraphics[width=0.96\columnwidth]{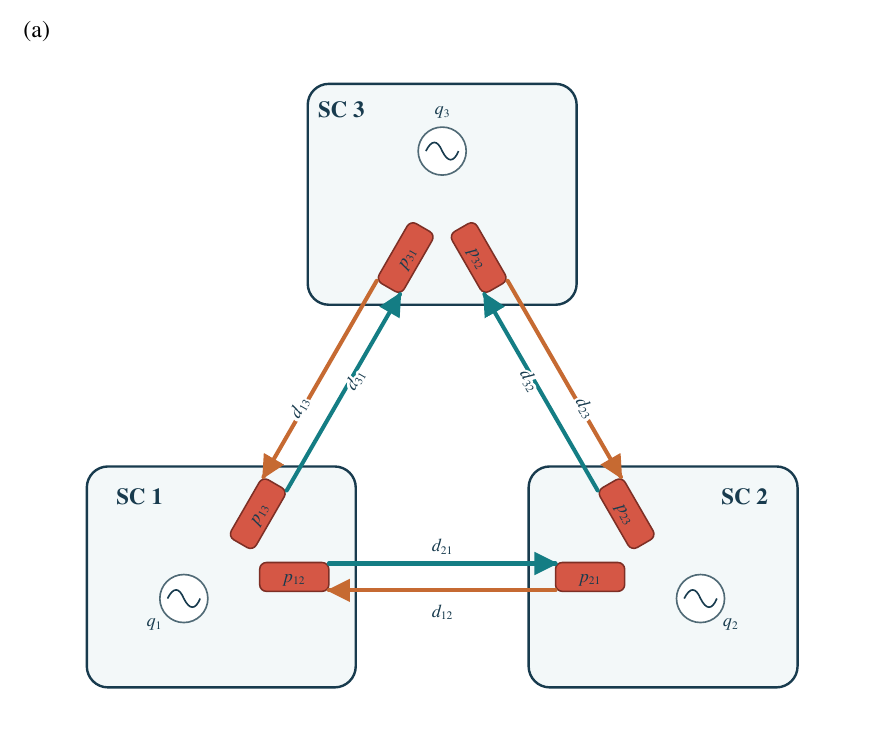}
\par\vspace{1mm}
\makebox[0.96\columnwidth][l]{{\fontsize{6}{7.2}\selectfont (b)}}
\par
\includegraphics[width=0.96\columnwidth,trim=55 55 55 60,clip]{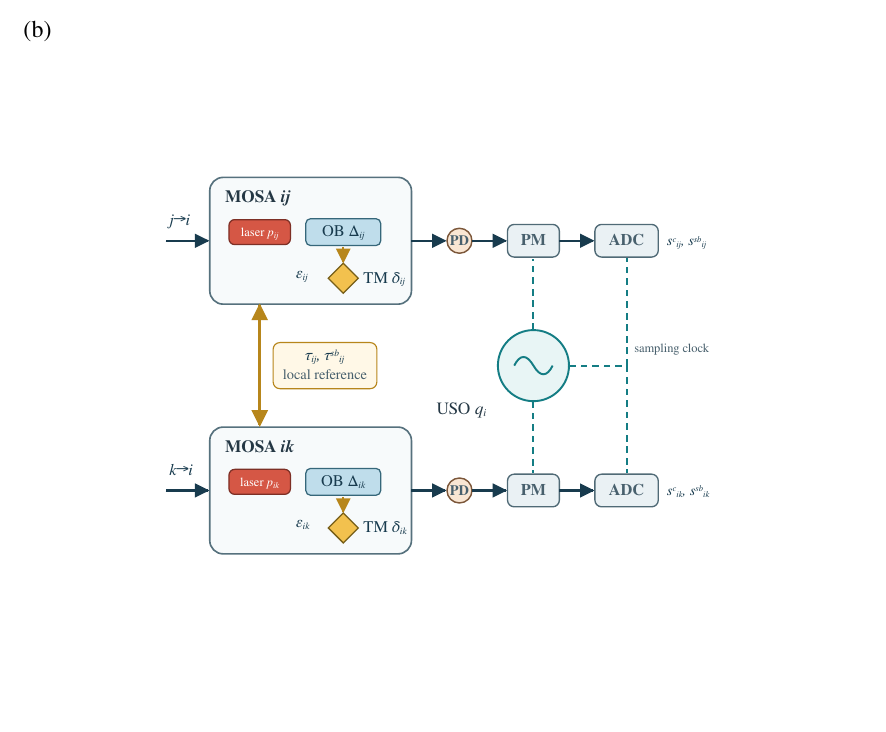}
\caption{\label{fig1}Measurement topology and index conventions.}
\end{inlinefigure}

We represent the detector as a directed network of three spacecraft.
An ordered pair $ij$ identifies the movable optical subassembly (MOSA) on spacecraft $i$ that faces spacecraft $j$.  The first index is local, and the second identifies the remote end of the arm.
When $i$, $j$, and $k$ occur together, they are distinct members of $\{1,2,3\}$.  Hence, $ij$ and $ik$ label the two MOSAs carried by spacecraft $i$.
This convention distinguishes the six lasers and fixes the orientation of every one-way measurement shown in Fig.~\ref{fig1}(a).

Each MOSA can provide carrier and upper-sideband phase measurements from the
interspacecraft, test-mass, and reference interferometers, giving six
readouts in the complete measurement model~\cite{tdi-clock-2021,unitmodelsim-2023}.
The present construction uses five of them.  The interspacecraft carrier
 measurement $s_{ij}^{c}$ compares the carrier received from MOSA $ji$ with
 the local carrier at MOSA $ij$ and contains the GW
response.  Its sideband counterpart $s_{ij}^{sb}$ transfers relative clock
and modulation information in addition to the optical phase.  The carrier
test-mass measurement $\varepsilon_{ij}$ measures the displacement of the
free-falling test mass relative to its optical bench.  The carrier reference
measurement $\tau_{ij}$ compares the two local lasers without a test-mass
reflection, while $\tau_{ij}^{sb}$ is its sideband counterpart.  The
test-mass sideband readout is part of the complete instrument model but does
not enter the calibration derived here and is therefore not assigned a
separate variable.  Figure~\ref{fig1}(b) summarizes the five readouts used
below.

The photodetector (PD), phasemeter (PM), and analog-to-digital converter
(ADC) form the readout chain shown in Fig.~\ref{fig1}(b).  The ADC on
spacecraft $i$ timestamps samples with its local USO time $t_i$, related to an
ideal coordinate time $t$ by $t_i=t+\delta t_i(t)$.
We denote the fluctuating timing error by $q_i\equiv\delta t_i$.
It has units of time, and its derivative $y_i\equiv\dot q_i=\delta f_i/f_0$ is the fractional USO frequency noise.  Its one-sided power spectral density (PSD) is denoted by $S_y(f)$, with $S_y(f)=(2\pi f)^2S_q(f)$.
Because all readouts are expressed as phase in cycles, a local heterodyne frequency multiplying $q_i$ produces the corresponding clock-induced phase error.
The nominal USO (or pilot-tone) frequency is denoted by $f_0$.
We retain the model only to first order and neglect products of clock jitter
with test-mass noise, optical-bench noise, optical-path noise, and other
secondary noises.

With these conventions, the five raw readouts are
\par
{\small
\begin{subequations}\label{origin}
\begin{align}
s_{ij}^c(t)={}&h_{ij}+D_{ij}p_{ji}-p_{ij}-a_{ij}q_i\notag\\*
&+\nu_{ji}D_{ij}\vec n_{ij}\!\cdot\!\vec\Delta_{ji}
 +\nu_{ji}\vec n_{ji}\!\cdot\!\vec\Delta_{ij}\notag\\*
&+N_{ij}^{\rm opt}+\int dt\,a_{ij},\label{sc}\\
s_{ij}^{sb}(t)={}&h_{ij}+D_{ij}p_{ji}-p_{ij}-c_{ij}q_i\notag\\*
&+(\nu_{ji}+\nu_{ji}^{m})D_{ij}\vec n_{ij}\!\cdot\!\vec\Delta_{ji}\notag\\*
&+(\nu_{ji}+\nu_{ji}^{m})\vec n_{ji}\!\cdot\!\vec\Delta_{ij}
 +N_{ij}^{\rm opt,sb}\notag\\*
&+\nu_{ji}^{m}D_{ij}(q_j+m_{ji})\notag\\* &-\nu_{ij}^{m}(q_i+m_{ij})
 +\int dt\,c_{ij},\label{sb}\\
\varepsilon_{ij}(t)={}&p_{ik}-p_{ij}-b_{ij}q_i+\mu_{ij}\notag\\*
&-2\nu_{ik}\vec n_{ji}\!\cdot\!(\vec\delta_{ij}-\vec\Delta_{ij})\notag\\* &
 +\int dt\,b_{ij},\label{test}\\
\tau_{ij}(t)={}&p_{ik}-p_{ij}-b_{ij}q_i+\mu_{ij}
 +\int dt\,b_{ij},\label{ref}\\
\tau_{ij}^{sb}(t)={}&p_{ik}-p_{ij}
 -(b_{ij}+\nu_{ik}^{m}-\nu_{ij}^{m})q_i+\mu_{ij}\notag\\*
&+\nu_{ik}^{m}(q_i+m_{ik})\notag\\* &-\nu_{ij}^{m}(q_i+m_{ij})\notag\\*
&+\int dt\,(b_{ij}+\nu_{ik}^{m}-\nu_{ij}^{m}).\label{refsb}
\end{align}
\end{subequations}
}
Here $h_{ij}$ is the GW-induced single-link optical phase (distinct from the dimensionless strain $h$ used later), and $p_{ij}$ is the phase noise of the laser on MOSA $ij$.
The vectors $\vec\delta_{ij}$ and $\vec\Delta_{ij}$ denote, respectively, test-mass displacement noise and optical-bench displacement noise.  The vector $\vec n_{ij}$ points from spacecraft $i$ toward spacecraft $j$ and projects those motions onto the arm.  Thus, $\vec n_{ji}$ follows the propagation direction of the link received at spacecraft $i$ from spacecraft $j$.
The terms $N_{ij}^{\rm opt}$ and $N_{ij}^{\rm opt,sb}$ collect optical-path and readout noise in the carrier and sideband interspacecraft measurements.
The quantity $\mu_{ij}$ is the phase noise of the local backlink or optical-fiber path shared by the test-mass and reference measurements.
The modulation-chain fluctuation $m_{ij}$ is written in the same timing-equivalent normalization as $q_i$, which is why the combinations $q_i+m_{ij}$ occur in the sideband measurements.
In the $c=1$ convention, $q_i$, $m_{ij}$, $d_{ij}$, $\vec\delta_{ij}$, and $\vec\Delta_{ij}$ are expressed as times, with the last two written as displacements divided by $c$.  The quantities $\nu_{ij}$, $\nu_{ij}^{m}$, $a_{ij}$, $b_{ij}$, and $c_{ij}$ are in hertz, and the measured quantities $s$, $h$, $p$, $N$, and $\mu$ are phases in cycles.

The optical carrier frequency on MOSA $ij$ is denoted by $\nu_{ij}$, and
$\nu_{ij}^{m}$ is its radio-frequency sideband offset.  The received phase of
a constant emitted frequency contains the argument $t-d^{\rm R}_{ij}(t)$.
Its derivative with respect to reception time therefore produces the factor
$1-\dot d^{\rm R}_{ij}$~\cite{tdi-clock-2021,unitmodelsim-2023}.  For light
emitted by MOSA $ji$ and received at MOSA $ij$, the reception-time dependence
of the frequencies is written explicitly as
\begin{align}
 \nu_{ji}(t)&\equiv[1-\dot d^{\rm R}_{ij}(t)]\nu_{ji},\notag\\
 \nu_{ji}^{m}(t)&\equiv[1-\dot d^{\rm R}_{ij}(t)]\nu_{ji}^{m}.
\end{align}
Here the explicit argument $t$ denotes the received frequency; the
frequency on the right-hand side is the constant emitted frequency.
Time arguments are otherwise suppressed in the measurement equations.
The coefficients multiplying the local clock error are the corresponding
heterodyne beat frequencies.  They are
$a_{ij}=\nu_{ji}-\nu_{ij}$ for the carrier interspacecraft measurement,
$b_{ij}=\nu_{ik}-\nu_{ij}$ for the local reference measurements, and
$c_{ij}=\nu_{ji}+\nu_{ji}^{m}-\nu_{ij}-\nu_{ij}^{m}$ for the sideband
interspacecraft measurement.
Here $\dot d^{\rm R}_{ij}$ is the time derivative of the reception-tagged
one-way light-travel time.  The integral terms in Eqs.~\eqref{origin} are
predictable phase ramps produced by the nominal beat frequencies, rather
than additional stochastic noise sources.  The notation
$\int dt\,a_{ij}$, for example, abbreviates
$\int_{t_0}^{t}a_{ij}(t')dt'$.  In the noise analysis these ramps are removed
with the deterministic heterodyne model.

Propagation from spacecraft $j$ to the receiver on spacecraft $i$ is represented by the retarded operator $D_{ij}$, which acts as $D_{ij}x(t)=x[t-d^{\rm R}_{ij}(t)]$.  Here $d^{\rm R}_{ij}(t)$ is the reception-tagged one-way light-travel time, and the speed of light is set to unity.
For a chain of links, concatenated subscripts denote successive delay operations, with the rightmost operator acting first.
Because the arm lengths vary, delay operators associated with different links need not commute.
The notation $D_{-ji}$ used below denotes the time-advance operator inverse to $D_{ij}$.  Its construction from the emission-tagged light-travel time is given in Sec.~\ref{sec:phase-locking-delays}.

\subsection{Roles of the reduced measurement streams}\label{section2.2}

Before an interferometric observable is assembled, the raw readouts in
Eqs.~\eqref{origin} are reorganized into the standard intermediary
measurements $\xi_{ij}$ and $\eta_{ij}$ together with independent clock
comparisons~\cite{tdi-03,tdi-clock-2021}.  No phase-locking constraint is
imposed in this subsection.

The intermediary measurement $\xi_{ij}$ augments $s_{ij}^{c}$ with
frequency-weighted test-mass--reference differences at the local end and at
the appropriately retimed remote end.  The difference
$\varepsilon-\tau$ removes the common local laser comparison and backlink
contribution while retaining the test-mass displacement relative to the
optical bench.  Thus, $\xi_{ij}$ refers the long-arm measurement to the
free-falling test masses without removing its GW response.  The subsequent
reference-measurement correction defines $\eta_{ij}$ and reduces the six MOSA
laser phases to three effective spacecraft laser phases.  With $i$, $j$, and
$k$ cyclically distinct, the clock-only parts before phase-locking are
\begin{subequations}\label{eta-noPL-clock}
\begin{align}
 \eta_{ij}^{q}={}&b_{jk}D_{ij}q_j-a_{ij}q_i,\\
 \eta_{ik}^{q}={}&-(b_{ij}+a_{ik})q_i,
\end{align}
\end{subequations}
where $b_{ij}=-b_{ik}$~\cite{tdi-clock-2021}.  The full $\eta$
measurements also contain the GW response, test-mass noise, optical-path
noise, and any modulation noise retained by the measurement model.

The sideband--carrier difference, after removal of its nominal beat phase and
normalization by the modulation frequency, provides the clock comparison
$r_{ij}$.  Its leading clock term in the ideal modulation limit is
\begin{align}\label{r-noPL-clock}
 r_{ij}^{q}=D_{ij}q_j-q_i.
\end{align}
The common carrier phase and GW response cancel in this difference.  The
reference pair $\tau_{ij}^{sb}$ and $\tau_{ij}$ provides the additional local
information used to remove differential modulation noise.  Section~\ref{section3}
shows how the Michelson phase-locking constraints reorganize these $\eta$ and
$r$ measurements.

\section{Measurement streams under phase-locking}\label{section3}

Section~\ref{section2} established the raw measurement model and summarized the reduction of measurements formed without phase-locking.
We now impose the Michelson phase-locking topology, relate its hardware constraints to the corresponding software elimination of redundant laser phases, and derive the resulting observable data streams.

\subsection{Phase-locking configuration}\label{section3.1}
Phase-locking establishes a directed optical-frequency network in which the
six lasers no longer represent independent phase processes.  It permits the
phase of a weak received field to be reproduced by a local transmitting laser
and keeps the planned heterodyne beats within the phasemeter band as the
orbital Doppler shifts evolve~\cite{phase-locking-tdi-2024,unitmodelsim-2023}.
We select the Michelson topology shown in Fig.~\ref{fig2}.  The laser on MOSA
12 supplies the primary reference.  The laser on MOSA 13 follows it through
the local reference interferometer, the lasers on MOSAs 21 and 31 follow the
fields received from spacecraft 1, and the remaining lasers follow the local
references on spacecraft 2 and 3.  This topology is one representative
choice from the phase-locking configurations analyzed in
Ref.~\cite{phase-locking-tdi-2024}.

\begin{inlinefigure}
\centering
\includegraphics[width=0.96\columnwidth]{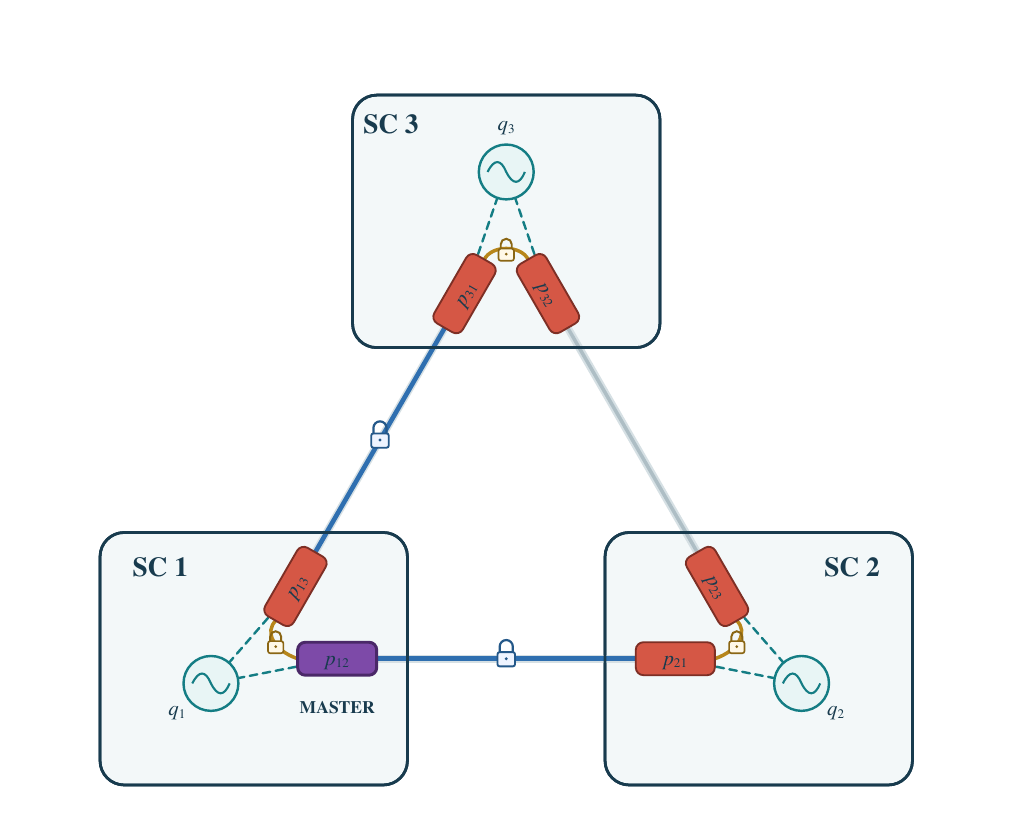}
\caption{\label{fig2}
Phase-locking configuration with the laser on MOSA 12 assigned as the primary
reference.}
\end{inlinefigure}

The local phase-locking relation between MOSAs 12 and 13 makes the corresponding
reference measurements equal in the ideal zero-error limit,
\begin{align}
 {\tau _{12}}(t) = {\tau _{13}}(t).
\end{align}
The two transmitted beams then establish the incoming-field locks at MOSAs 21
and 31.  The local reference interferometers propagate those phase references
to MOSAs 23 and 32.  In the notation of Eq.~\eqref{origin}, the four ideal carrier
and local-reference phase-locking error signals satisfy
\begin{align}
{s_{21}^{c}}(t) =& 0,\notag\\
{s_{31}^{c}}(t) =& 0,\notag\\
{\tau _{23}}(t) =& {\tau _{21}}(t),\notag\\
{\tau _{31}}(t) =& {\tau _{32}}(t).
\end{align}
Restricting these relations to the laser-phase-noise sector reduces the six
laser phase noises to delayed copies of the MOSA-12 process as follows.
\begin{align}
{p_{13}}(t) =& {p_{12}}(t) \equiv p_1(t),\notag\\
{p_{23}}(t) =& {p_{21}}(t) = {D_{21}}p_1(t),\notag\\
{p_{32}}(t) =& {p_{31}}(t) = {D_{31}}p_1(t).
\end{align}
Finite phase-locking residuals are outside the ideal model used below.
\subsection{Relation between algebraic laser-noise reduction and ideal phase-locking}\label{section3.2}

The software reduction from the raw readouts to the $\xi$ and $\eta$
measurements, summarized in Sec.~\ref{section2.2}, has the same laser-noise
transfer structure as the ideal, zero-error limit of the phase-locking
architecture.  This correspondence does not make the complete observables
identical.  USO fluctuations, modulation noise, and any finite phase-locking
residual must still be propagated separately.

For the observable data stream from the left-hand MOSA on spacecraft $j$, it
is required that the data stream be consistent with the clock reference
output, i.e., ${p_{ji}}(t)-{p_{jk}}(t)+\int_{}^t dt\,{b_{jk}}
\to\int_{}^{\bar t}dt\,{b_{jk}}$.  Since the clock on spacecraft $j$ is not
perfect, one finds
${p_{ji}}(\bar t)-{p_{jk}}(\bar t)-{b_{jk}}{q_j}\to0$.
Phase-locking is therefore a phase-tracking process referenced to the local
clock on spacecraft $j$, which contains the fluctuation $q_j$.  When
${p_{ji}}(t)$ is eliminated, the relation becomes
\begin{align}
{p_{ji}}(t) + {b_{jk}}t = {p_{jk}}(t) + {b_{jk}}\int_{}^t {(1 + {{\dot q}_j})dt}.
\end{align}
Here ${\dot q}_j$ is the first derivative of $q_j$ with respect to time.  The
bias signal $\int_{}^t dt\,{b_{jk}}$ is delayed and then sampled at spacecraft
$i$, producing the clock term
$-(1-{\dot d_{ij}}){b_{jk}}{q_i}\simeq-{b_{jk}}{q_i}$.  The term
${b_{jk}}{q_j}$ introduced by phase-locking becomes
${b_{jk}}D_{ij}q_j$ after propagation.  Their sum is the additional clock
contribution in the $ij$-oriented measurement,
${b_{jk}}(D_{ij}q_j-q_i)={b_{jk}}r_{ij}^{q}$.  Suppressing this contribution
therefore amounts to subtracting the clock transfer introduced by the
intra-spacecraft phase-locking operation.

For the complementary data stream from the left-hand MOSA on spacecraft $i$,
the relation
${p_{ik}}(t)-{p_{ij}}(t)+\int_{}^t dt\,{b_{ij}}
\to\int_{}^{\bar t}dt\,{b_{ij}}$ can be written as
${p_{ik}}(t)+{b_{ij}}t={p_{ij}}(t)+{b_{ij}}
\int_{}^t(1+{\dot q}_i)dt$.  In this orientation, the clock term
${b_{ij}}\int_{}^t{\dot q}_i dt$ associated with phase-locking cancels the
sampling contribution of the phase ramp ${b_{ij}}t$.  No additional
phase-locking clock term therefore enters $\eta_{ik}$.

Within the ideal zero-error approximation, the laser-noise constraints
imposed by hardware phase-locking can be represented by the corresponding
algebraic elimination applied to measurements formed without phase-locking.
For example, eliminating
${D_{ij}}{p_{ji}}(t\circ t_i)={D_{ij}}{p_{ji}}(t-q_i)
\approx {D_{ij}}{p_{ji}}(t)$ and ${p_{ik}}(t)$ from the carrier measurement
introduces the following correction terms.
\begingroup
\setlength{\abovedisplayskip}{4pt}
\setlength{\belowdisplayskip}{4pt}
\setlength{\jot}{1pt}
\begin{equation}\label{brintor}
\begin{aligned}
&-D_{ij}\frac{\tau_{jk}(t\circ t_j)-\tau_{ji}(t\circ t_j)}{2}\\
&\quad=-D_{ij}[p_{ji}(t-q_j)-p_{jk}(t-q_j)]\\
&\qquad-D_{ij}\int^{t-q_j}dt\,b_{jk}\\
&\quad\approx-D_{ij}[p_{ji}(t)-p_{jk}(t)]
\\ &\qquad-b_{jk}(t-q_j-d_{ij})\\
&\quad\approx-\int^{t-d_{ij}}dt\,b_{jk}
\\ &\qquad-D_{ij}[p_{ji}(t)-p_{jk}(t)]+b_{jk}D_{ij}q_j,
\end{aligned}
\end{equation}
and
\begin{equation}\label{0intor}
\begin{aligned}
&\frac{\tau_{ij}(t\circ t_i)-\tau_{ik}(t\circ t_i)}{2}\\
&\quad=p_{ik}(t-q_i)-p_{ij}(t-q_i)
 +\int^{t-q_i}dt\,b_{ij}\\
&\quad\approx p_{ik}(t)-p_{ij}(t)+b_{ij}(t-q_i)\\
&\quad\approx\int^t dt\,b_{ij}+p_{ik}(t)-p_{ij}(t)-b_{ij}q_i.
\end{aligned}
\end{equation}
\endgroup
The notation $t\circ t_i$ indicates that $t$ is evaluated as a function of
the local clock time $t_i$.  In Eq.~\eqref{brintor}, local sampling of the
delayed bias phase supplies $-{b_{jk}}q_i$.  Together with
${b_{jk}}D_{ij}q_j$, it gives the intra-spacecraft phase-locking contribution
${b_{jk}}r_{ij}^{q}$.  In Eq.~\eqref{0intor}, the sampling term
${b_{ij}}q_i$ cancels the explicit term $-{b_{ij}}q_i$.  These approximations
retain only first-order clock couplings.

\subsection{Reduced data streams under phase-locking}\label{section3.3}
For the Michelson phase-locking topology in Fig.~\ref{fig2}, $s_{21}^c(t)\to0$ and $s_{31}^c(t)\to0$.  The four reduced streams are defined by
\begin{subequations}\label{defintils}
\begin{equation}
\tilde s_1=s_{12}^c+D_{12}s_{21}^c.
\end{equation}
\begin{equation}
\begin{aligned}
\tilde s_{1'}={}&s_{13}^c+D_{13}s_{31}^c\\
&+(1-D_{13}D_{31})\tau_1.
\end{aligned}
\end{equation}
\begin{equation}
\begin{aligned}
\tilde s_2={}&s_{23}^c+D_{23}s_{31}^c-s_{21}^c\\
&-D_{23}\tau_3-D_{23}D_{31}\tau_1-\tau_2.
\end{aligned}
\end{equation}
\begin{equation}
\begin{aligned}
\tilde s_{3'}={}&s_{32}^c+D_{32}s_{21}^c-s_{31}^c\\
&+\tau_3+D_{32}\tau_2+D_{31}\tau_1.
\end{aligned}
\end{equation}
\end{subequations}
\par
\end{multicols}
In Eqs.~\eqref{defintils}, $\tau_i$ denote the three intra-spacecraft phase-locking null streams,
\begin{align}\label{defintau}
{\tau _i}(t) = \frac{{{\tau _{ij}}(t) - {\tau _{ik}}(t)}}{2} = {p_{ik}}(t) - {p_{ij}}(t) - {b_{ij}}{q_i}(t) \to 0.
\end{align}

The arrow denotes ideal nulling of the phase-locking error signal.  It neither
sets $q_i$ to zero nor assumes synchronization of the three onboard clocks.
Finite phase-locking residuals are neglected in this study.
Substituting Eqs.~\eqref{origin} into Eq.~\eqref{defintau} and then Eqs.~\eqref{defintils}, we obtain
\par
{\small
\begin{subequations}
\begin{align}
{{\tilde s}_1}(t) =& {h_{12}} + {D_{12}}{h_{21}} + {D_{12}}{D_{21}}{p_1} - {p_1} + {\nu _{21}}[{D_{12}}{{\vec n}_{12}} \cdot {{\vec \Delta }_{21}} + {{\vec n}_{21}} \cdot {{\vec \Delta }_{12}}] \notag\\
 &+ {D_{12}}{\nu _{12}}[{D_{21}}{{\vec n}_{21}} \cdot {{\vec \Delta }_{12}} + {{\vec n}_{12}} \cdot {{\vec \Delta }_{21}}]\notag\\
+& N_{12}^{opt} + {D_{12}}N_{21}^{opt} + \int {dt} {a_{12}} + {D_{12}}\int {dt} {a_{21}} - {a_{12}}{q_1} - {D_{12}}{a_{21}}{q_2},\label{tildes1}\\
{{\tilde s}_{1'}}(t) =& {h_{13}} + {D_{13}}{h_{31}} + {D_{13}}{D_{31}}{p_1} - {p_1} + {\nu _{31}}[{D_{13}}{{\vec n}_{13}} \cdot {{\vec \Delta }_{31}} + {{\vec n}_{31}} \cdot {{\vec \Delta }_{13}}] \notag\\
 &+ {D_{13}}{\nu _{13}}[{D_{31}}{{\vec n}_{31}} \cdot {{\vec \Delta }_{13}} + {{\vec n}_{13}} \cdot {{\vec \Delta }_{31}}]\notag\\
 +& N_{13}^{opt} + {D_{13}}N_{31}^{opt} + \int {dt} {a_{13}} + {D_{13}}\int {dt} {a_{31}} - ({a_{13}} + {b_{12}}){q_1} - {D_{13}}{a_{31}}{q_3} + {D_{13}}{D_{31}}{b_{12}}{q_1},\label{tildes1p}\\
{{\tilde s}_2}(t) =& {h_{23}} + {D_{23}}{h_{31}} - {h_{21}} + ({D_{23}}{D_{31}} - {D_{21}}){p_1} + {\nu _{32}}[{D_{23}}{{\vec n}_{23}} \cdot {{\vec \Delta }_{32}} + {{\vec n}_{32}} \cdot {{\vec \Delta }_{23}}] \notag\\
 &+ {D_{23}}{\nu _{13}}[{D_{31}}{{\vec n}_{31}} \cdot {{\vec \Delta }_{13}} + {{\vec n}_{13}} \cdot {{\vec \Delta }_{31}}]\notag\\
 -& {\nu _{12}}[{D_{21}}{{\vec n}_{21}} \cdot {{\vec \Delta }_{12}} + {{\vec n}_{12}} \cdot {{\vec \Delta }_{21}}] + N_{23}^{opt} + {D_{23}}N_{31}^{opt} - N_{21}^{opt} + \int {dt} {a_{23}} + {D_{23}}\int {dt} {a_{31}} - \int {dt} {a_{21}}\notag\\
 -& ({a_{23}} - {a_{21}} - {b_{23}}){q_2} - {D_{23}}({a_{31}} - {b_{31}}){q_3} + {D_{23}}{D_{31}}{b_{12}}{q_1},\label{tildes2}\\
 {{\tilde s}_{3'}}(t) =& {h_{32}} + {D_{32}}{h_{21}} - {h_{31}} + ({D_{32}}{D_{21}} - {D_{31}}){p_1} + {\nu _{23}}[{D_{32}}{{\vec n}_{32}} \cdot {{\vec \Delta }_{23}} + {{\vec n}_{23}} \cdot {{\vec \Delta }_{32}}] \notag\\
 &+ {D_{32}}{\nu _{12}}[{D_{21}}{{\vec n}_{21}} \cdot {{\vec \Delta }_{12}} + {{\vec n}_{12}} \cdot {{\vec \Delta }_{21}}]\notag\\
 -& {\nu _{13}}[{D_{31}}{{\vec n}_{31}} \cdot {{\vec \Delta }_{13}} + {{\vec n}_{13}} \cdot {{\vec \Delta }_{31}}] + N_{32}^{opt} + {D_{32}}N_{21}^{opt} - N_{31}^{opt} + \int {dt} {a_{32}} + {D_{32}}\int {dt} {a_{21}} - \int {dt} {a_{31}}\notag\\
 -& ({a_{32}} - {a_{31}} + {b_{31}}){q_3} - {D_{32}}({a_{21}} + {b_{23}}){q_2} - {D_{31}}{b_{12}}{q_1}.
\end{align}\label{tildes}
\end{subequations}
}

The sideband--carrier difference provides an interspacecraft clock comparison.
To first order in the receiver-clock error, the two orientations are
\par
\begin{multicols}{2}

\begin{subequations}\label{denr}
\begin{align}
{r_{ij}}(t) ={}& \frac{s_{ij}^{sb}(t)-s_{ij}^c(t)}{\nu_{ji}^m}
-\frac{1}{\nu_{ji}^m}\int dt\,(\nu_{ji}^m-\nu_{ij}^m) \notag\\
 \approx{}& {D_{ij}}q_j-(1-\dot d_{ij})q_i-d_{ij}+{D_{ij}}\Delta m_j,\\*
{r_{ik}}(t) ={}& \frac{s_{ik}^{sb}(t)-s_{ik}^c(t)}{\nu_{ki}^m}
-\frac{1}{\nu_{ki}^m}\int dt\,(\nu_{ki}^m-\nu_{ik}^m) \notag\\
 \approx{}& {D_{ik}}q_k-(1-\dot d_{ik})q_i-d_{ik}-\frac{{\nu _{ik}^m}}{{\nu_{ki}^m}}\Delta m_i,
\end{align}
\end{subequations}
where $\Delta m_i$ is the differential modulation noise obtained from the
reference carrier and sideband measurements,
\begin{align}\label{Delatm}
\Delta m_i \equiv{}&
 \frac{\tau _{ij}^{sb}-\tau _{ij}-\int dt\,(\nu_{ik}^m-\nu_{ij}^m)}{2\nu_{ik}^m}\notag\\
&-\frac{\tau _{ik}^{sb}-\tau _{ik}+\int dt\,(\nu_{ik}^m-\nu_{ij}^m)}{2\nu_{ik}^m}\notag\\
={}&m_{ik}-\frac{\nu_{ij}^m}{\nu_{ik}^m}m_{ij}.
\end{align}
Neglecting the products $\dot d_{ij}q_i$ at the accuracy used below, the
phase-locking topology gives
\begin{subequations}
\begin{align}
{r_{21}}(t) \approx{}& ({D_{21}}{q_1} - {q_2}) - {d_{21}} - \frac{{\nu _{21}^m}}{{\nu_{12}^m}}\Delta {m_2},\\
{r_{31}}(t) \approx{}& ({D_{31}}{q_1} - {q_3}) - {d_{31}} + {D_{31}}\Delta {m_1}.
\end{align}\label{denrphas}
\end{subequations}
\par
\end{multicols}
Under phase-locking, the carrier streams are further reduced as
\begin{subequations}
\begin{align}
{s_1}(t) \equiv& {\tilde s_1}(t) - {D_{12}}{a_{21}}{r_{21}}(t),\label{s1}\\
{s_{1'}}(t) \equiv& {\tilde s_{1'}}(t) - {D_{13}}{a_{31}}{r_{31}}(t),\label{s1p}\\
{s_2}(t) \equiv& {\tilde s_2}(t) - ({a_{23}} - {a_{21}} - {b_{23}}){r_{21}}(t) - {D_{23}}({a_{31}} - {b_{31}}){r_{31}}(t),\label{s2}\\
{s_{3'}}(t) \equiv& {\tilde s_{3'}}(t) - ({a_{32}} - {a_{31}} + {b_{31}}){r_{31}}(t) - {D_{32}}({a_{21}} + {b_{23}}){r_{21}}(t).
\end{align}\label{s}
\end{subequations}

Substituting Eqs.~\eqref{tildes} and \eqref{denrphas} into Eqs.~\eqref{s}, we obtain
\par
{\small
\begin{subequations}
\begin{align}
{s_1}(t) =& {h_{12}} + {D_{12}}{h_{21}} + {D_{12}}{D_{21}}{p_1} - {p_1} + {\nu _{21}}[{D_{12}}{{\vec n}_{12}} \cdot {{\vec \Delta }_{21}} + {{\vec n}_{21}} \cdot {{\vec \Delta }_{12}}] \notag\\
 &+ {D_{12}}{\nu _{12}}[{D_{21}}{{\vec n}_{21}} \cdot {{\vec \Delta }_{12}} + {{\vec n}_{12}} \cdot {{\vec \Delta }_{21}}]\notag\\
+& N_{12}^{opt} + {D_{12}}N_{21}^{opt} + \int {dt} {a_{12}} + {D_{12}}\int {dt} {a_{21}} - {a_{12}}{q_1} - {D_{12}}{a_{21}}[{D_{21}}{q_1} - {d_{21}} - \frac{{\nu _{21}^m}}{{\nu_{12}^m}}\Delta {m_2}],\label{exes1}\\
{s_{1'}}(t) =& {h_{13}} + {D_{13}}{h_{31}} + {D_{13}}{D_{31}}{p_1} - {p_1} + {\nu _{31}}[{D_{13}}{{\vec n}_{13}} \cdot {{\vec \Delta }_{31}} + {{\vec n}_{31}} \cdot {{\vec \Delta }_{13}}] \notag\\
 &+ {D_{13}}{\nu _{13}}[{D_{31}}{{\vec n}_{31}} \cdot {{\vec \Delta }_{13}} + {{\vec n}_{13}} \cdot {{\vec \Delta }_{31}}]\notag\\
 +& N_{13}^{opt} + {D_{13}}N_{31}^{opt} + \int {dt} {a_{13}} + {D_{13}}\int {dt} {a_{31}} - ({a_{13}} + {b_{12}}){q_1} - {D_{13}}{a_{31}}[{D_{31}}{q_1} - {d_{31}} + {D_{31}}\Delta {m_1}]\notag\\ &+ {D_{13}}{D_{31}}{b_{12}}{q_1},\label{exes1p}\\
{s_2}(t) =& {h_{23}} + {D_{23}}{h_{31}} - {h_{21}} + ({D_{23}}{D_{31}} - {D_{21}}){p_1} + {\nu _{32}}[{D_{23}}{{\vec n}_{23}} \cdot {{\vec \Delta }_{32}} + {{\vec n}_{32}} \cdot {{\vec \Delta }_{23}}] \notag\\
 &+ {D_{23}}{\nu _{13}}[{D_{31}}{{\vec n}_{31}} \cdot {{\vec \Delta }_{13}} + {{\vec n}_{13}} \cdot {{\vec \Delta }_{31}}]\notag\\
 -& {\nu _{12}}[{D_{21}}{{\vec n}_{21}} \cdot {{\vec \Delta }_{12}} + {{\vec n}_{12}} \cdot {{\vec \Delta }_{21}}] + N_{23}^{opt} + {D_{23}}N_{31}^{opt} - N_{21}^{opt} + \int {dt} {a_{23}} + {D_{23}}\int {dt} {a_{31}} - \int {dt} {a_{21}}\notag\\
 -& ({a_{23}} - {a_{21}} - {b_{23}})[{D_{21}}{q_1} - {d_{21}} - \frac{{\nu _{21}^m}}{{\nu_{12}^m}}\Delta {m_2}] - {D_{23}}({a_{31}} - {b_{31}})[{D_{31}}{q_1} - {d_{31}} + {D_{31}}\Delta {m_1}] + {D_{23}}{D_{31}}{b_{12}}{q_1},\label{exes2}\\
{s_{3'}}(t) = &{h_{32}} + {D_{32}}{h_{21}} - {h_{31}} + ({D_{32}}{D_{21}} - {D_{31}}){p_1} + {\nu _{23}}[{D_{32}}{{\vec n}_{32}} \cdot {{\vec \Delta }_{23}} + {{\vec n}_{23}} \cdot {{\vec \Delta }_{32}}] \notag\\
 &+ {D_{32}}{\nu _{12}}[{D_{21}}{{\vec n}_{21}} \cdot {{\vec \Delta }_{12}} + {{\vec n}_{12}} \cdot {{\vec \Delta }_{21}}]\notag\\
 -& {\nu _{13}}[{D_{31}}{{\vec n}_{31}} \cdot {{\vec \Delta }_{13}} + {{\vec n}_{13}} \cdot {{\vec \Delta }_{31}}] + N_{32}^{opt} + {D_{32}}N_{21}^{opt} - N_{31}^{opt} + \int {dt} {a_{32}} + {D_{32}}\int {dt} {a_{21}} - \int {dt} {a_{31}}\notag\\
 -& ({a_{32}} - {a_{31}} + {b_{31}})[{D_{31}}{q_1} - {d_{31}} + {D_{31}}\Delta {m_1}] - {D_{32}}({a_{21}} + {b_{23}})[{D_{21}}{q_1} - {d_{21}} - \frac{{\nu _{21}^m}}{{\nu_{12}^m}}\Delta {m_2}] - {D_{31}}{b_{12}}{q_1}.
\end{align}\label{exes}
\end{subequations}
}

The following combinations remove optical-bench and differential modulation
noise from the phase-locking measurements.
\par
{\small
\begin{subequations}
\begin{align}
{\tilde \eta _1}(t) \equiv{}& {s_1}(t) - \frac{\nu _{21}}{2\nu _{23}}{D_{12}}{\varepsilon _{21}}(t) - \frac{\nu _{21}}{2\nu _{13}}{\varepsilon _{12}}(t) - \frac{\nu _{12}}{2\nu _{13}}{D_{12}}{D_{21}}{\varepsilon _{12}}(t) \notag\\
 &- \frac{\nu _{12}}{2\nu _{23}}{D_{12}}{\varepsilon _{21}}(t) - {D_{12}}{a_{21}}\frac{\nu _{21}^m}{\nu_{12}^m}\Delta {m_2},\label{tildeeta1}\\
{\tilde \eta _{1'}}(t) \equiv{}& {s_{1'}}(t) - \frac{\nu _{31}}{2\nu _{32}}{D_{13}}{\varepsilon _{31}}(t) - \frac{\nu _{31}}{2\nu _{12}}{\varepsilon _{13}}(t) - \frac{\nu _{13}}{2\nu _{12}}{D_{13}}{D_{31}}{\varepsilon _{13}}(t) \notag\\
 &- \frac{\nu _{13}}{2\nu _{32}}{D_{13}}{\varepsilon _{31}}(t) + {D_{13}}{a_{31}}{D_{31}}\Delta {m_1}(t),\label{tildeeta1p}\\
{{\tilde \eta }_2}(t) \equiv{}& {s_2}(t) - \frac{\nu _{32}}{2\nu _{31}}{D_{23}}{\varepsilon _{32}}(t) - \frac{\nu _{32}}{2\nu _{21}}{\varepsilon _{23}}(t) - \frac{\nu _{13}}{2\nu _{12}}{D_{23}}{D_{31}}{\varepsilon _{13}}(t) \notag\\
 &- \frac{\nu _{13}}{2\nu _{32}}{D_{23}}{\varepsilon _{31}}(t) + \frac{\nu _{12}}{2\nu _{13}}{D_{21}}{\varepsilon _{12}}(t) + \frac{\nu _{12}}{2\nu _{23}}{\varepsilon _{21}}(t)\notag\\
 -& ({a_{23}} - {a_{21}} - {b_{23}})\frac{\nu _{21}^m}{\nu_{12}^m}\Delta {m_2} + {D_{23}}({a_{31}} - {b_{31}}){D_{31}}\Delta {m_1}(t),\label{tildeeta2}\\
{{\tilde \eta }_{3'}}(t) \equiv{}& {s_{3'}}(t) - \frac{\nu _{23}}{2\nu _{21}}{D_{32}}{\varepsilon _{23}}(t) - \frac{\nu _{23}}{2\nu _{31}}{\varepsilon _{32}}(t) - \frac{\nu _{12}}{2\nu _{13}}{D_{32}}{D_{21}}{\varepsilon _{12}}(t) \notag\\
 &- \frac{\nu _{12}}{2\nu _{23}}{D_{32}}{\varepsilon _{21}}(t) + \frac{\nu _{13}}{2\nu _{12}}{D_{31}}{\varepsilon _{13}}(t) + \frac{\nu _{13}}{2\nu _{32}}{\varepsilon _{31}}(t)\notag\\
 +& ({a_{32}} - {a_{31}} + {b_{31}}){D_{31}}\Delta {m_1}(t) - {D_{32}}({a_{21}} + {b_{23}})\frac{\nu _{21}^m}{\nu_{12}^m}\Delta {m_2}(t).
\end{align}\label{exetildeeta}
\end{subequations}
}

By inserting Eqs.~\eqref{exes}, \eqref{test} and \eqref{Delatm} into Eqs.~\eqref{exetildeeta}, we derive
\par
{\small
\begin{subequations}
\begin{align}
{{\tilde \eta }_1}(t) =& {h_{12}} + {D_{12}}{h_{21}} + {D_{12}}{D_{21}}{p_1} - {p_1} + {\nu _{21}}[{D_{12}}{{\vec n}_{12}} \cdot {{\vec \delta }_{21}} + {{\vec n}_{21}} \cdot {{\vec \delta }_{12}}] \notag\\
 &+ {D_{12}}{\nu _{12}}[{D_{21}}{{\vec n}_{21}} \cdot {{\vec \delta }_{12}} + {{\vec n}_{12}} \cdot {{\vec \delta }_{21}}]\notag\\
 +& N_{12}^{opt} + {D_{12}}N_{21}^{opt} + \int {dt} {a_{12}} + {D_{12}}\int {dt} {a_{21}} - {a_{12}}{q_1} - {D_{12}}{a_{21}}\left( {{D_{21}}{q_1} - {d_{21}}} \right),\label{exetildeeta1}\\
{{\tilde \eta }_{1'}}(t) =& {h_{13}} + {D_{13}}{h_{31}} + {D_{13}}{D_{31}}{p_1} - {p_1} + {\nu _{31}}[{D_{13}}{{\vec n}_{13}} \cdot {{\vec \delta }_{31}} + {{\vec n}_{31}} \cdot {{\vec \delta }_{13}}] \notag\\
 &+ {D_{13}}{\nu _{13}}[{D_{31}}{{\vec n}_{31}} \cdot {{\vec \delta }_{13}} + {{\vec n}_{13}} \cdot {{\vec \delta }_{31}}]\notag\\
 +& N_{13}^{opt} + {D_{13}}N_{31}^{opt} + \int {dt} {a_{13}} + {D_{13}}\int {dt} {a_{31}} - ({a_{13}} + {b_{12}}){q_1} - {D_{13}}{a_{31}}[{D_{31}}{q_1} - {d_{31}}] + {D_{13}}{D_{31}}{b_{12}}{q_1},\label{exetildeeta1p}\\
{{\tilde \eta }_2}(t) =& {h_{23}} + {D_{23}}{h_{31}} - {h_{21}} + ({D_{23}}{D_{31}} - {D_{21}}){p_1} + {\nu _{32}}[{D_{23}}{{\vec n}_{23}} \cdot {{\vec \delta }_{32}} + {{\vec n}_{32}} \cdot {{\vec \delta }_{23}}] \notag\\
 &+ {D_{23}}{\nu _{13}}[{D_{31}}{{\vec n}_{31}} \cdot {{\vec \delta }_{13}} + {{\vec n}_{13}} \cdot {{\vec \delta }_{31}}]\notag\\
 -& {\nu _{12}}[{D_{21}}{{\vec n}_{21}} \cdot {{\vec \delta }_{12}} + {{\vec n}_{12}} \cdot {{\vec \delta }_{21}}] + N_{23}^{opt} + {D_{23}}N_{31}^{opt} - N_{21}^{opt} + \int {dt} {a_{23}} + {D_{23}}\int {dt} {a_{31}} - \int {dt} {a_{21}}\notag\\
 -& ({a_{23}} - {a_{21}} - {b_{23}})[{D_{21}}{q_1} - {d_{21}}] - {D_{23}}({a_{31}} - {b_{31}})[{D_{31}}{q_1} - {d_{31}}] + {D_{23}}{D_{31}}{b_{12}}{q_1},\label{exetildeeta2}\\
{{\tilde \eta }_{3'}}(t) =& {h_{32}} + {D_{32}}{h_{21}} - {h_{31}} + ({D_{32}}{D_{21}} - {D_{31}}){p_1}(t) + {\nu _{23}}[{D_{32}}{{\vec n}_{32}} \cdot {{\vec \delta }_{23}} + {{\vec n}_{23}} \cdot {{\vec \delta }_{32}}] \notag\\
 &+ {D_{32}}{\nu _{12}}[{D_{21}}{{\vec n}_{21}} \cdot {{\vec \delta }_{12}} + {{\vec n}_{12}} \cdot {{\vec \delta }_{21}}]\notag\\
 -& {\nu _{13}}[{D_{31}}{{\vec n}_{31}} \cdot {{\vec \delta }_{13}} + {{\vec n}_{13}} \cdot {{\vec \delta }_{31}}] + N_{32}^{opt} + {D_{32}}N_{21}^{opt} - N_{31}^{opt} + \int {dt} {a_{32}} + {D_{32}}\int {dt} {a_{21}} - \int {dt} {a_{31}}\notag\\
 -& ({a_{32}} - {a_{31}} + {b_{31}})[{D_{31}}{q_1} - {d_{31}}] - {D_{32}}({a_{21}} + {b_{23}})[{D_{21}}{q_1} - {d_{21}}] - {D_{31}}{b_{12}}{q_1}.
\end{align}\label{fullexetildeeta}
\end{subequations}
}

We form four composite sideband measurements $r_i$ under phase-locking,
\par
\begin{multicols}{2}

\begin{subequations}
\begin{align}
{r_1}(t) \equiv& {r_{12}}(t) + {D_{12}}{r_{21}}(t),\label{defr1}\\
{r_{1'}}(t) \equiv& {r_{13}}(t) + {D_{13}}{r_{31}}(t),\label{defr1p}\\
{r_2}(t) \equiv& {r_{23}}(t) + {D_{23}}{r_{31}}(t) - {r_{21}}(t),\label{defr2}\\
{r_{3'}}(t) \equiv&{r_{32}}(t) + {D_{32}}{r_{21}}(t) - {r_{31}}(t).
\end{align}\label{defr}
\end{subequations}
Substituting Eq.~\eqref{denr} into Eq.~\eqref{defr} gives
\begin{subequations}\label{exedefr}
\begin{equation}\label{exedefr1}
\begin{aligned}
r_1={}&(D_{12}D_{21}-1)q_1-d_{12}-D_{12}d_{21}\\
&+D_{12}\left(1-\frac{\nu_{21}^{m}}{\nu_{12}^{m}}\right)\Delta m_2.
\end{aligned}
\end{equation}
\begin{equation}\label{exedefr1p}
\begin{aligned}
r_{1'}={}&(D_{13}D_{31}-1)q_1-d_{13}-D_{13}d_{31}\\
&+\left(D_{13}D_{31}-\frac{\nu_{13}^{m}}{\nu_{31}^{m}}\right)\Delta m_1.
\end{aligned}
\end{equation}
\begin{equation}\label{exedefr2}
\begin{aligned}
r_2={}&(D_{23}D_{31}-D_{21})q_1-d_{23}-D_{23}d_{31}+d_{21}\\
&+D_{23}\Delta m_3+D_{23}D_{31}\Delta m_1
 +\frac{\nu_{21}^{m}}{\nu_{12}^{m}}\Delta m_2.
\end{aligned}
\end{equation}
\begin{equation}
\begin{aligned}
r_{3'}={}&(D_{32}D_{21}-D_{31})q_1-d_{32}-D_{32}d_{21}+d_{31}\\
&-\frac{\nu_{32}^{m}}{\nu_{23}^{m}}\Delta m_3
 -D_{32}\frac{\nu_{21}^{m}}{\nu_{12}^{m}}\Delta m_2-D_{31}\Delta m_1.
\end{aligned}
\end{equation}
\end{subequations}
After the differential modulation terms are removed and the known ranging
terms are retained explicitly, the cleaned sideband measurements are
\begin{subequations}\label{tildedefr}
\begin{equation}\label{tildedefr1}
\begin{aligned}
\widetilde r_1\equiv{}&r_1-D_{12}\left(1-\frac{\nu_{21}^{m}}{\nu_{12}^{m}}\right)\Delta m_2\\
\approx{}&(D_{12}D_{21}-1)q_1-d_{12}-D_{12}d_{21}.
\end{aligned}
\end{equation}
\begin{equation}\label{tildedefr1p}
\begin{aligned}
\widetilde r_{1'}\equiv{}&r_{1'}-
 \left(D_{13}D_{31}-\frac{\nu_{13}^{m}}{\nu_{31}^{m}}\right)\Delta m_1\\
\approx{}&(D_{13}D_{31}-1)q_1-d_{13}-D_{13}d_{31}.
\end{aligned}
\end{equation}
\begin{equation}\label{tildedefr2}
\begin{aligned}
\widetilde r_2\equiv{}&D_{-12}\left(r_2-D_{23}\Delta m_3-D_{23}D_{31}\Delta m_1\right.\\
&\left.\hspace{9mm}-\frac{\nu_{21}^{m}}{\nu_{12}^{m}}\Delta m_2\right)\\
\simeq{}&(D_{-12}D_{23}D_{31}-1)q_1\\
&-D_{-12}(d_{23}+D_{23}d_{31}-d_{21}).
\end{aligned}
\end{equation}
\begin{equation}
\begin{aligned}
\widetilde r_{3'}\equiv{}&D_{-13}\left(r_{3'}+
 \frac{\nu_{32}^{m}}{\nu_{23}^{m}}\Delta m_3\right.\\
&\left.\hspace{5mm}+D_{32}\frac{\nu_{21}^{m}}{\nu_{12}^{m}}\Delta m_2
 +D_{31}\Delta m_1\right)\\
\simeq{}&(D_{-13}D_{32}D_{21}-1)q_1\\
&-D_{-13}(d_{32}+D_{32}d_{21}-d_{31}).
\end{aligned}
\end{equation}
\end{subequations}
The four phase-locking measurements used in the subsequent construction are
\begin{subequations}\label{etadef}
\begin{align}
\eta_1&\equiv\widetilde\eta_1+a_{21}\widetilde r_1,\label{eta1}\\
\eta_{1'}&\equiv\widetilde\eta_{1'}+(a_{31}-b_{12})\widetilde r_{1'},\label{eta1p}\\
\eta_2&\equiv D_{-12}\widetilde\eta_2
 +(a_{31}-b_{31}-b_{12})\widetilde r_2,\label{eta2}\\
\eta_{3'}&\equiv D_{-13}\widetilde\eta_{3'}
 +(a_{21}+b_{23})\widetilde r_{3'}.
\end{align}
\end{subequations}
\par
\end{multicols}
Substituting Eqs.~\eqref{fullexetildeeta} and~\eqref{tildedefr} into Eqs.~\eqref{etadef} gives the combined phase-locking streams,
\par
{\small
\begin{subequations}\label{fulletadef}
\begin{equation}\label{exeeta1}
\begin{aligned}
\eta_1={}&h_{12}+D_{12}h_{21}+(D_{12}D_{21}-1)p_1\\
&+\nu_{21}[D_{12}\vec n_{12}\!\cdot\!\vec\delta_{21}
 +\vec n_{21}\!\cdot\!\vec\delta_{12}]
 +D_{12}\nu_{12}[D_{21}\vec n_{21}\!\cdot\!\vec\delta_{12}
 +\vec n_{12}\!\cdot\!\vec\delta_{21}]\\
&+N_{12}^{\rm opt}+D_{12}N_{21}^{\rm opt}
 +\int dt\,a_{12}+D_{12}\int dt\,a_{21}
 -(a_{12}+a_{21})q_1-a_{21}d_{12}.
\end{aligned}
\end{equation}
\begin{equation}\label{exeeta1p}
\begin{aligned}
\eta_{1'}={}&h_{13}+D_{13}h_{31}+(D_{13}D_{31}-1)p_1\\
&+\nu_{31}[D_{13}\vec n_{13}\!\cdot\!\vec\delta_{31}
 +\vec n_{31}\!\cdot\!\vec\delta_{13}]
 +D_{13}\nu_{13}[D_{31}\vec n_{31}\!\cdot\!\vec\delta_{13}
 +\vec n_{13}\!\cdot\!\vec\delta_{31}]\\
&+N_{13}^{\rm opt}+D_{13}N_{31}^{\rm opt}
 +\int dt\,a_{13}+D_{13}\int dt\,a_{31}
 -(a_{13}+a_{31})q_1-a_{31}d_{13}+b_{12}(d_{13}+D_{13}d_{31}).
\end{aligned}
\end{equation}
\begin{equation}\label{exeeta2}
\resizebox{0.96\textwidth}{!}{$\displaystyle
\begin{aligned}
\eta_2={}&D_{-12}(h_{23}+D_{23}h_{31}-h_{21})
 +(D_{-12}D_{23}D_{31}-1)p_1
 +D_{-12}\!\bigl\{\nu_{32}[D_{23}\vec n_{23}\!\cdot\!\vec\delta_{32}
 +\vec n_{32}\!\cdot\!\vec\delta_{23}]
 +D_{23}\nu_{13}[D_{31}\vec n_{31}\!\cdot\!\vec\delta_{13}
 +\vec n_{13}\!\cdot\!\vec\delta_{31}]\\
&\qquad-\nu_{12}[D_{21}\vec n_{21}\!\cdot\!\vec\delta_{12}
 +\vec n_{12}\!\cdot\!\vec\delta_{21}]\bigr\}
 +D_{-12}\!\left(N_{23}^{\rm opt}+D_{23}N_{31}^{\rm opt}-N_{21}^{\rm opt}
 +\int dt\,a_{23}+D_{23}\int dt\,a_{31}-\int dt\,a_{21}\right)\\
&\quad-(a_{23}-b_{31}+a_{31}-b_{12}-a_{21}-b_{23})(q_1-D_{-12}d_{21})
 -(a_{31}-b_{31}-b_{12})D_{-12}d_{23}+b_{12}D_{-12}D_{23}d_{31}.
\end{aligned}$}
\end{equation}
\begin{equation}
\begin{aligned}
\eta_{3'}={}&D_{-13}(h_{32}+D_{32}h_{21}-h_{31})
 +(D_{-13}D_{32}D_{21}-1)p_1\\
&+\nu_{23}[D_{-13}D_{32}\vec n_{32}\!\cdot\!\vec\delta_{23}
 +D_{-13}\vec n_{23}\!\cdot\!\vec\delta_{32}]
 +D_{-13}D_{32}\nu_{12}[D_{21}\vec n_{21}\!\cdot\!\vec\delta_{12}
 +\vec n_{12}\!\cdot\!\vec\delta_{21}]\\
&-\nu_{13}[\vec n_{31}\!\cdot\!\vec\delta_{13}
 +D_{-13}\vec n_{13}\!\cdot\!\vec\delta_{31}]
 +D_{-13}\left(N_{32}^{\rm opt}+D_{32}N_{21}^{\rm opt}-N_{31}^{\rm opt}
 +\int dt\,a_{32}+D_{32}\int dt\,a_{21}-\int dt\,a_{31}\right)\\
&-(a_{32}+b_{31}-a_{31}+a_{21}+b_{23}+b_{12})(q_1-D_{-13}d_{31})
 -b_{12}D_{-13}d_{31}-(a_{21}+b_{23})D_{-13}d_{32}.
\end{aligned}
\end{equation}
\end{subequations}
}

Equations~\eqref{fulletadef} provide the phase-locking data streams used below to construct the clock-noise calibration for geometric TDI.
\par
\begin{multicols}{2}

\section{Geometric TDI under phase-locking}\label{section4}

This section develops the delay representation needed by the phase-locking
measurements.  We first define the topology-dependent composite delays and
their inverse retimings, and then express a two-branch geometric observable in
a single-master basis.

\subsection{Equivalent time-delay operators under phase-locking}
\label{sec:phase-locking-delays}
\par
\end{multicols}
Within the ideal zero-phase-locking-error approximation adopted in Sec.~\ref{section3.2}, the slave-laser fluctuations are not independent noise degrees of freedom. In the laser-noise sector, they are delayed replicas of the master-laser phase noise $p_1$. Consequently, the laser-noise contributions to the four phase-locking data streams in Eqs.~\eqref{fulletadef} are different ordered transfers of the same stochastic process.
\begin{subequations}
\begin{align}
{\eta _{\rm{1}}}(t) =& {D_{12}}{D_{21}}{p_1} - {p_1} = \left( {{D_{12}}{D_{21}} - {\rm{1}}} \right){p_1},\\
{\eta _{{\rm{1'}}}}(t) =& {D_{13}}{D_{31}}{p_1} - {p_1} = \left( {{D_{13}}{D_{31}} - {\rm{1}}} \right){p_1},\\
{\eta _2}(t) =& {D_{ - 12}}{D_{23}}{D_{31}}{p_1} - {p_1} = \left( {{D_{ - 12}}{D_{23}}{D_{31}} - 1} \right){p_1},\\
{\eta _{3'}}(t) =& {D_{ - 13}}{D_{32}}{D_{21}}{p_1} - {p_1} = \left( {{D_{ - 13}}{D_{32}}{D_{21}} - 1} \right){p_1}.
\end{align}\label{etap}
\end{subequations}

Equation~\eqref{etap} displays the phase-locking-specific reduction. The four streams differ through the ordered propagation chains applied to $p_1$, rather than through independent laser noises associated with different optical benches. We therefore collect each complete phase-locking propagation history into a composite operator.
\par
\begin{multicols}{2}

\begin{inlinefigure}
\centering
\includegraphics[width=0.90\columnwidth]{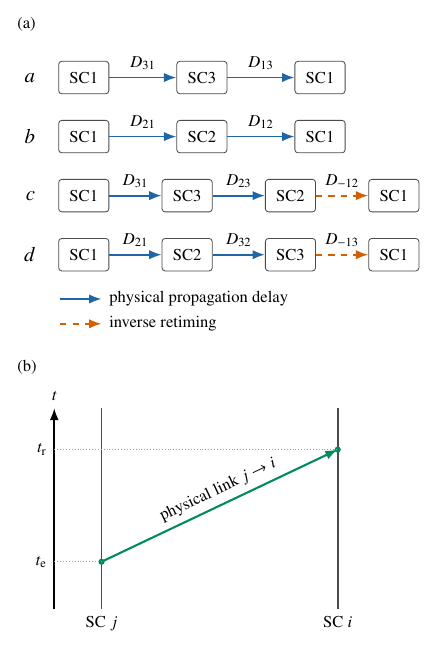}
\caption{\label{fig3}
Construction of the phase-locking delay operators.  Panel (a) shows the four
ordered composite histories.  Solid arrows denote physical propagation
delays, and dashed arrows denote inverse retiming.  Panel (b) shows the
reception-tagged and emission-tagged descriptions of the same one-way optical
link.  The emission-tagged light-travel time is required by the inverse
factors in $\eta_2$ and $\eta_{3'}$.}
\end{inlinefigure}

\begin{subequations}
\begin{align}
a \equiv& {D_{13}}{D_{31}} = {D_{131}},\\
b \equiv& {D_{12}}{D_{21}} = {D_{121}},\\
c \equiv& {D_{ - 12}}{D_{23}}{D_{31}},\\
d \equiv& {D_{ - 13}}{D_{32}}{D_{21}},
\end{align}\label{equiva op}
\end{subequations}
Figure~\ref{fig3} illustrates the four composite histories in
Eq.~\eqref{equiva op} and the reception- and emission-tagged light times
required by the inverse delays.

The operators $a$ and $b$ describe round-trip histories based at spacecraft 1
through spacecraft 3 and spacecraft 2, respectively.  The operators $c$ and $d$ contain an inverse
one-way delay that retimes the corresponding phase-locking streams to the
common master-laser reference.  The four positive composite histories used in
their numerical construction are
\begin{equation}\label{positive-composite-delays}
\begin{aligned}
 D_7&=D_{23}D_{31}, & D_8&=D_{13}D_{31},\\
 D_9&=D_{32}D_{21}, & D_{10}&=D_{12}D_{21},
\end{aligned}
\end{equation}
so that $a=D_8$, $b=D_{10}$, $c=D_{-12}D_7$, and $d=D_{-13}D_9$.
This replaces the six elementary one-way delay histories used without
phase-locking by four topology-dependent composite histories.

For a time-dependent constellation, the reception-tagged and emission-tagged
light-travel times must be distinguished.  Let $d^{\rm R}_{ij}(t_{\rm r})$
denote the light-travel time for the physical link from spacecraft $j$ to
spacecraft $i$ when the reception time is known.  Let
$d^{\rm E}_{ij}(t_{\rm e})$ denote the same light-travel time when the emission
time is known.  They satisfy
\begin{equation}\label{emission-reception-times}
\begin{aligned}
 t_{\rm r}&=t_{\rm e}+d^{\rm E}_{ij}(t_{\rm e}),\\
 d^{\rm E}_{ij}(t_{\rm e})&=d^{\rm R}_{ij}(t_{\rm r}),\\
 d^{\rm E}_{ij}(t)&=d^{\rm R}_{ij}\!\left[t+d^{\rm E}_{ij}(t)\right].
\end{aligned}
\end{equation}
Accordingly, the retarded operator and its inverse act as
\begin{equation}\label{retarded-inverse-actions}
\begin{aligned}
 D_{ij}x(t)&=x\!\left[t-d^{\rm R}_{ij}(t)\right],\\
 D_{-ji}x(t)&=x\!\left[t+d^{\rm E}_{ij}(t)\right].
\end{aligned}
\end{equation}
In particular, $D_{-12}=D_{21}^{-1}$ and $D_{-13}=D_{31}^{-1}$.
The implicit equation for $d^{\rm E}_{ij}$ is solved iteratively from the
reception-tagged light-travel times.  For fractional-frequency data, the two
retimings also contain the factors $1-\dot d^{\rm R}_{ij}$ and
$1+\dot d^{\rm E}_{ij}$, respectively.  The inverse factors in $\eta_2$ and
$\eta_{3'}$ therefore require $d^{\rm E}_{21}$ and $d^{\rm E}_{31}$ in
addition to the reception-tagged light-travel times.  An inverse delay is a
data-retiming operation and does not represent backward propagation of light.

\subsection{Single-source representation of geometric TDI}

Phase-locking changes the laser-noise content assigned to a geometric TDI path. Without phase-locking, one-way measurements along a synthesized path involve laser phases belonging to different spacecraft or optical benches. Under the phase-locking constraints used here, those phases reduce to delayed copies of one master process, which we henceforth denote by $p\equiv p_1$. Each virtual branch can therefore be represented by an ordered sequence formed from the composite operators in Eq.~\eqref{equiva op} and their inverses.

\par
\end{multicols}
Consider two such operator sequences associated with synthesized branches that share a terminal reference event. Because both sequences act on the same master-laser phase process, their unmatched contribution is a scalar operator difference applied to $p$.
\begin{align}\label{georesidual}
\mathrm{TDI}_\mathrm{res}^\mathrm{geo}=\left ({D}_{\pm i_1} {D}_{\pm i_2} \ldots{D}_{\pm i_n} -{D}_{\pm i_1^{\prime}}{D}_{\pm i_2^{\prime}} \ldots {D}_{\pm i_n^{\prime}}\right ) p.
\end{align}

Here, each $i_k$ or $i_k^{\prime}$ selects one of the composite operators $a$, $b$, $c$, and $d$. We use $D_{+i}=D_i$ for the corresponding delay and $D_{-i}=D_i^{-1}$ for its inverse.  Equivalently, the signed operator alphabet is $\{a,b,c,d,\bar a,\bar b,\bar c,\bar d\}$. The signs are part of the operator index rather than algebraic prefactors. Factor ordering is retained because delays associated with time-dependent arms do not generally commute. Writing the two ordered sequences in compact product notation, Eq.~\eqref{georesidual} reads
\begin{align}\label{geoexpression}
\mathrm{TDI}_\mathrm{res}^\mathrm{geo}= \left( {\prod\limits_{k = 1}^n {{D}_{ \pm {i_k}}}  - \prod\limits_{k = 1}^n {{D}_{ \pm {i_k^{\prime}}}} } \right){p}.
\end{align}
\par
\begin{multicols}{2}

The modified first-generation Michelson observable $X_1$ directly illustrates this single-source construction. In the composite phase-locking basis, its two branch sequences are $ab$ and $ba$. Since both act on the same master-laser phase $p_1$, the residual takes the commutator form
\begin{align}\label{Xresdiual}
\mathrm{TDI}_{X_1}^p = (ab -ba)p_1 .
\end{align}
Thus, Eq.~\eqref{Xresdiual} is $[a,b]p_1$. It vanishes when the effective operators commute, as in the constant-arm limit, whereas for time-dependent arms it measures the mismatch produced by the different orderings. Figure~\ref{fig4} traces both ordered branches to their common terminal event on spacecraft 1.

\begin{inlinefigure}
\centering
\includegraphics[width=0.60\columnwidth]{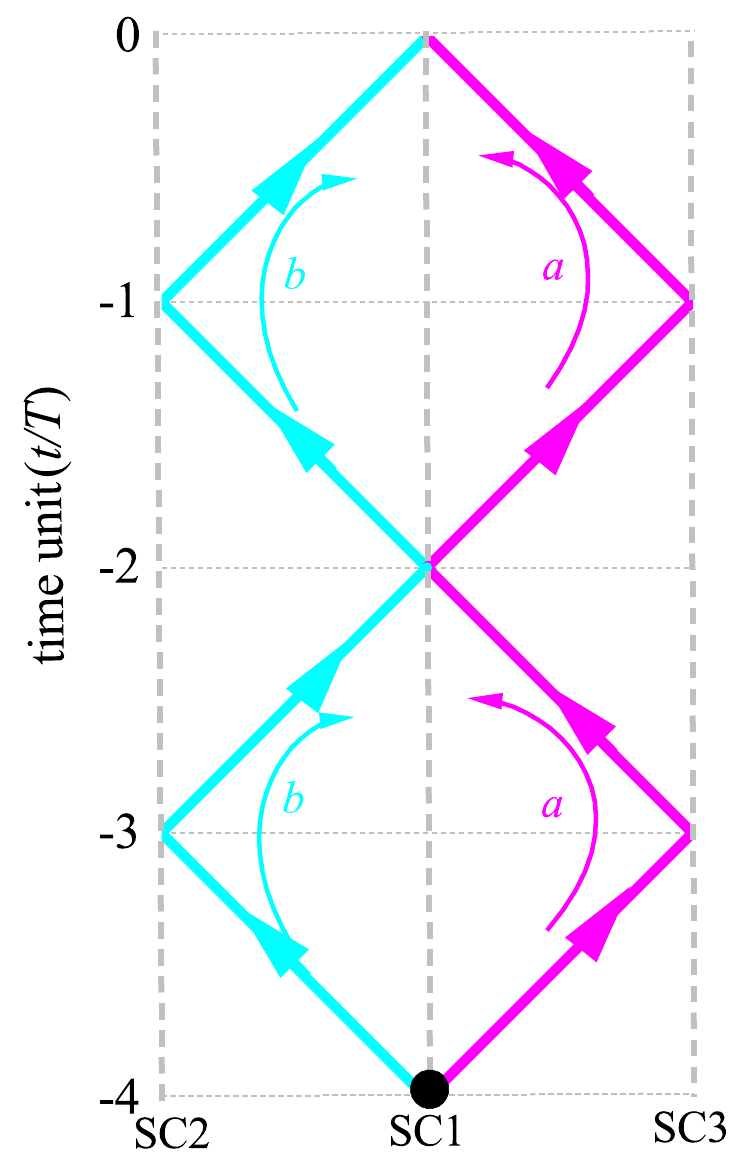}
\caption{\label{fig4} 
Composite-operator space-time representation of the modified first-generation Michelson observable $X_1$ under phase-locking. The two synthesized branches realize the ordered sequences $ab$ and $ba$. The black markers denote their initial events, while their common terminal event on spacecraft 1 defines the reference time $t=0$.}
\end{inlinefigure}

We next express the residual from an arbitrary ordered operator sequence in terms of the available phase-locking data streams. Although the telescoping step below is an algebraic identity, its building blocks here are composite streams driven by one master-laser noise rather than the one-way streams without phase-locking, which involve distinct laser phases. Retaining only the laser-noise part, each positive-index stream compares a transformed copy of the common phase $p$ with the reference copy.
\begin{align}\label{plp}
{\eta _i} = {D_i}p - p.
\end{align}
Here, $i\in\{a,b,c,d\}$ labels a complete composite chain, $p=p_1$, and $\eta_i$ denotes only the laser-noise contribution to the corresponding full data stream. To include an inverse chain without introducing another laser-noise degree of freedom, we define its negative-index stream by
\begin{align}\label{newplp}
{\eta _{ - i}} \equiv {D_{ - i}}p - p =  - {D_{ - i}}({D_i}p - p) =  - {D_{ - i}}{\eta _i}.
\end{align}
This identity uses $D_{-i}D_i=I$.  The negative-index quantity is generated from the positive-index stream by inverse retiming and is not an additional physical readout. Its two terms still contain the same master process $p$.

\par
\end{multicols}
Now consider an arbitrary ordered sequence ${D}_{\pm i_1}{D}_{\pm i_2}\cdots {D}_{\pm i_n}$. Inserting the common reference phase between successive partial products decomposes the difference between the final and initial copies of $p$ into a telescoping sequence of phase-locking streams.
\par
{\small
\begin{align}\label{intermeres}
{D_{ \pm {i_1}}}{D_{ \pm {i_2}}}...{D_{ \pm {i_n}}}p - p =& \left( {\prod\limits_{k = 1}^n {{D_{ \pm {i_k}}}} } \right)p - p\notag\\
 =& {D_{ \pm {i_1}... \pm {i_{n - 1}}}}({D_{ \pm {i_n}}}p - p) + {D_{ \pm {i_1}... \pm {i_{n - 2}}}}({D_{ \pm {i_{n - 1}}}}p - p)\notag\\ &+ ... + {D_{ \pm {i_1}}}({D_{ \pm {i_2}}}p - p) + ({D_{ \pm {i_1}}}p - p)\notag\\
 =& {D_{ \pm {i_1}.... \pm {i_{n - 1}}}}{\eta _{ \pm {i_n}}} + {D_{ \pm {i_1}... \pm {i_{n - 2}}}}{\eta _{ \pm {i_{n - 1}}}} + ... + {D_{ \pm {i_1}}}{\eta _{ \pm {i_2}}} + {\eta _{ \pm {i_1}}}.
 \end{align}
}

Equation~\eqref{intermeres} preserves the operator order and therefore does not assume commuting delays. In the phase-locking representation, every term is a transfer of the same $p$, so the expansion requires no spacecraft- or bench-dependent laser labels.  The construction without phase-locking must instead track the sequence of different laser phases along the path. Applying the construction separately to the two branches in Eq.~\eqref{georesidual} gives their representation in terms of the available $\eta_{\pm i}$ streams.
\par
{\small
\begin{align}\label{interetaLASER}
 {\rm {TDI}^{geo}} =& ({\eta _{ \pm {i_1}}} + {D_{ \pm {i_1}}}{\eta _{ \pm {i_2}}} + ... + {D_{ \pm {i_1}}}{D_{ \pm {i_2}}}...{D_{ \pm {i_{n - 1}}}}{\eta _{ \pm {i_n}}}) \notag\\
 &- ({\eta _{ \pm {i_1^{\prime}}}} + {D_{ \pm {i_1^{\prime}}}}{\eta _{ \pm {i_2^{\prime}}}} + ... + {D_{ \pm {i_1^{\prime}}}}{D_{ \pm {i_2^{\prime}}}}...{D_{ \pm {i_{n - 1}^{\prime}}}}{\eta _{ \pm {i_n^{\prime}}}})\notag\\
 =& \sum\limits_{k = 1}^n {\left\{ {\left( {\prod\limits_{s = 1}^{k - 1} {{D_{ \pm {i_s}}}} } \right){\eta _{ \pm {i_k}}} - \left( {\prod\limits_{s = 1}^{k - 1} {{D_{ \pm {i_s^{\prime}}}}} } \right){\eta _{ \pm {i_k^{\prime}}}}} \right\}}.
 \end{align}
}

Equation~\eqref{interetaLASER} is therefore a single-master representation of a two-branch geometric TDI observable. Its phase-locking-specific content is the reduction of the laser-noise sector to one stochastic process and the use of the composite alphabet $\{a,b,c,d\}$ to encode the accessible phase-locking propagation histories. This representation also supplies the laser-noise transfer structure used for the clock-noise construction in Sec.~\ref{section5}.
\par
\begin{multicols}{2}

\section{Residual clock noise and its reduction algorithm}\label{section5}

The phase-locking representation developed in Sec.~\ref{section4} cancels the master-laser fluctuation by comparing two ordered sequences of composite delay operators. Its clock sector requires a separate treatment because laser phase-locking does not synchronize the onboard USOs. In the reduced streams of Sec.~\ref{section3}, the remote-clock differences are transferred to sideband-derived clock observables, while the remaining unobserved clock contribution is referred to the spacecraft-1 process $q\equiv q_1$. We use this reference-clock representation and the same composite-operator sequences to construct an observable calibration template for geometric TDI.

\subsection{Clock-noise calibration in the phase-locking reference basis}\label{section5.1}

Let the unprimed and primed branches be specified by two ordered sequences of signed composite symbols. Each $i_k$ or $i_k^{\prime}$ selects one of $a$, $b$, $c$, and $d$, and a negative sign selects the inverse retiming operation. As a zeroth-order light-time check, a common endpoint requires equality of the accumulated signed propagation times.
\begin{align}\label{dequalarm}
\sum_{k=1}^{n}\left(d_{\pm i_k}-d_{\pm i_k^{\prime}}\right)=0.
\end{align} 
Here, $d_{-i}=-d_i$ is used as scalar bookkeeping, and $d_{\pm i}$ is the signed propagation time of a complete phase-locking composite history rather than that of an elementary one-way link. For time-dependent arms, Eq.~\eqref{dequalarm} is necessary but not by itself sufficient. Exact closure must be verified from the ordered delay operators, or equivalently from coincidence of the geometric endpoints. The calibration construction begins only after that dynamic closure test and the clock-coupling test below are imposed.

The single $q$ used below is an algebraic reference representation, not a claim that the three physical USOs have identical fluctuations. The carrier-sideband combinations constructed in Sec.~\ref{section3.3} measure delayed inter-spacecraft clock differences, including $D_{21}q_1-q_2$ and $D_{31}q_1-q_3$, and thereby transfer the remote-clock information to clock observables. Accordingly, the clock-only part of a positive composite stream can be written as $\eta_i^q=-\xi_i q$, where the scalar $\xi_i$ is the effective beat-note coupling read from the corresponding phase-locking stream in Eqs.~\eqref{fulletadef}. This coefficient should not be confused with the TDI intermediary measurement $\xi_{ij}$ introduced in Sec.~\ref{section2}. The data and clock-observable assignments associated with Eq.~\eqref{equiva op} are
\begin{align*}
a&\leftrightarrow(\eta_{1'},\widetilde r_{1'}),&
b&\leftrightarrow(\eta_1,\widetilde r_1),\\
c&\leftrightarrow(\eta_2,\widetilde r_2),&
d&\leftrightarrow(\eta_{3'},\widetilde r_{3'}).
\end{align*}
Their phase-locking-dependent clock couplings are
\begin{align*}
 \xi_a={}&a_{13}+a_{31},\\
 \xi_b={}&a_{12}+a_{21},\\
 \xi_c={}&a_{23}-b_{31}+a_{31}-b_{12}-a_{21}-b_{23},\\
 \xi_d={}&a_{32}+b_{31}-a_{31}+a_{21}+b_{23}+b_{12}.
\end{align*}
These four coefficients, rather than a generic one-way-link coefficient table, are the quantities used when a signed sequence is converted into a clock template.

The compact factorized expressions below adopt a frozen-coupling approximation.  Each $\xi_i$ is taken to be constant, or to vary negligibly over every composite retiming and prefix in the signed operator sequence. Under this approximation, define ${\xi _{ - i}} =  - {\xi _i}$ and $q_{-i}\equiv D_{-i}q$ for an inverse symbol. The relation $\eta _{ - i}^q = {D_{ - i}}[{\xi _i}q] =  - {\xi _{ - i}}{D_{ - i}}q \equiv  - {\xi _{ - i}}{q_{ - i}}$ then holds. Together with $q_{+i}\equiv q$, both orientations are covered by $\eta _{ \pm i}^q =  - {\xi _{ \pm i}}{q_{ \pm i}}$. If the beat-note couplings are treated as time dependent, every delay must instead act on the complete product.  In particular, one retains $D_{-i}[\xi_i(t)q(t)]$ and uses $\xi_{-i}(t)=-D_{-i}\xi_i(t)$. For bookkeeping, we denote the completed two-branch clock quantity by

\begin{align}\label{clockres}
\widetilde {\rm TDI}_{\rm geo}^q={}&
-\sum_{k = 1}^n \left\{
 \left( \prod_{s = 1}^{k - 1} {D}_{\pm i_s} \right)
 \xi_{\pm i_k}q_{\pm i_k}\right.\notag\\
&\left.\hspace{17mm}
- \left( \prod_{s = 1}^{k - 1} {D}_{\pm i_s^{\prime}} \right)
 \xi_{\pm i_k^{\prime}}q_{\pm i_k^{\prime}} \right\}\notag\\
&+\sum_{k = 1}^n
 \left(\xi_{\pm i_k}-\xi_{\pm i_k^{\prime}}\right)q.
\end{align}
The second sum completes every transported clock sample with the same undelayed reference $q$. Thus, Eq.~\eqref{clockres} coincides with the physical clock residual only when the selected phase-locking scheme and the two operator sequences satisfy the coefficient-closure condition $\sum_k(\xi_{\pm i_k}-\xi_{\pm i_k^{\prime}})=0$.  Otherwise the added term is a nonzero remainder. We treat this condition as an explicit admissibility check, complementary to the dynamic propagation closure. Once it holds, the residual can be organized entirely into differences between retimed copies of the reference clock.

\par
\end{multicols}
After the modulation and ranging corrections described in Sec.~\ref{section3.3}, the sideband measurements provide the required clock-difference building blocks. In the equations below, $r_i$ denotes the clock-only projection of the cleaned observable $\widetilde r_i$ in Eq.~\eqref{tildedefr}, rather than the uncorrected raw sideband ratio, which also contains modulation, ranging, and readout terms. For two consecutive positive composite symbols, the ideal clock-only quantity is
\begin{align}\label{r}
 {r_{ + {i_k}, + {i_{k + 1}}}} = {D_{{i_k}}}q - q,
 \end{align}
In this notation, $r_{+i,+j}\equiv r_i$.  The second entry records the orientation of the following composite step and does not introduce another clock process. The other three sign assignments are synthesized retimings of the same measured clock comparison.
\begin{subequations}
\begin{align}
{r_{ - {i_k}, + {i_{k + 1}}}} =& {D_{ - {i_k}}}q - {q_{ - {i_k}}} = {D_{ - {i_k}}}q - {D_{ - {i_k}}}q = 0,\\
{r_{{i_k}, - {i_{k + 1}}}} =& {D_{{i_k}}}{q_{ - {i_{k + 1}}}} - {q_{{i_k}}}
= {D_{{i_k}}}{D_{ - {i_{k + 1}}}}q - q \notag\\
=& - {D_{{i_k}}}{D_{ - {i_{k + 1}}}}({D_{{i_{k + 1}}}}q - q)
+ {D_{{i_k}}}q - q \notag\\
=& - {D_{{i_k}}}{D_{ - {i_{k + 1}}}}{r_{{i_{k + 1}}}} + {r_{{i_k}}},\\
{r_{ - {i_k}, - {i_{k + 1}}}} =& {D_{ - {i_k}}}{q_{ - {i_{k + 1}}}} - {q_{ - {i_k}}}
= {D_{ - {i_k}}}{D_{ - {i_{k + 1}}}}q - {D_{ - {i_k}}}q \notag\\
=& - {D_{ - {i_k}}}{D_{ - {i_{k + 1}}}}({D_{{i_{k + 1}}}}q - q) \notag\\
=& - {D_{ - {i_k}}}{D_{ - {i_{k + 1}}}}{r_{{i_{k + 1}}}}.
\end{align}\label{newr}
\end{subequations}

Equation~\eqref{newr} requires no additional phasemeter readout. The first orientation vanishes identically, and the remaining two are obtained from positive-orientation clock observables by applying the indicated composite retimings. These identities use $D_{-i}D_i=I$ to the accuracy of the adopted delay model. Inverse operators in this dictionary are post-processing operations.  They do not represent clocks or optical signals evolving physically backward in time.

Consider the prefix ending at the $k$th symbol of either branch. The difference between its propagated reference-clock history and the common starting value can be decomposed into increments at adjacent composite steps.
\begin{align}\label{kq}
&\left( {\prod\limits_{s = 1}^{k - 1} {{D_{ \pm {i_s}}}} } \right){q_{ \pm {i_k}}} - q\notag\\
 =& {D_{ \pm {i_1}}}{D_{ \pm {i_2}}}...{D_{ \pm {i_{k - 1}}}}{q_{ \pm {i_k}}} - q\notag\\
 =& {D_{ \pm {i_1}... \pm {i_{k - 2}}}}
 ({D_{ \pm {i_{k - 1}}}}{q_{ \pm {i_k}}} - {q_{ \pm {i_{k - 1}}}})\notag\\
 &+ {D_{ \pm {i_1}... \pm {i_{k - 3}}}}
 ({D_{ \pm {i_{k - 2}}}}{q_{ \pm {i_{k - 1}}}} - {q_{ \pm {i_{k - 2}}}}) + ...\notag\\
 &+ {D_{ \pm {i_1}}}({D_{ \pm {i_2}}}{q_{ \pm {i_3}}} - {q_{ \pm {i_2}}})
 + ({D_{ \pm {i_1}}}{q_{ \pm {i_2}}} - {q_{ \pm {i_1}}})\notag\\
 =& {r_{ \pm {i_1}, \pm {i_2}}} + {D_{\pm {i_1}}}{r_{ \pm {i_2}, \pm {i_3}}} + ... + {D_{ \pm {i_1}}}{D_{ \pm {i_2}}}...{D_{ \pm {i_{k - 2}}}}{r_{ \pm {i_{k - 1}}, \pm {i_k}}}\notag\\
 =& \sum\limits_{m = 1}^{k - 1} {\left( {\prod\limits_{s = 1}^{m - 1} {{D_{ \pm {i_s}}}} } \right){r_{ \pm {i_m}, \pm {i_{m + 1}}}}}.
\end{align}

Equation~\eqref{kq} is an ordered telescoping identity and does not assume commuting delays. The signed-link algebra is standard in geometric clock-noise calibration~\cite{clock-2023-Yang}.  Its distinct role here is to translate a prefix of a phase-locking composite sequence into the corresponding sideband clock comparisons. Whenever an inverse symbol occurs, Eq.~\eqref{newr} selects the realizable retiming needed for the associated adjacent pair.

Weighting the prefix reconstruction at position $k$ by the corresponding effective coupling and applying it to both branches produces the clock-calibration template
\par
{\small\begin{align}\label{all}
{\rm TDI}^q ={}&
-\sum_{k = 1}^n\sum_{m = 1}^{k - 1}
\left\{\xi_{\pm i_k}
\left(\prod_{s = 1}^{m - 1}D_{\pm i_s}\right)
r_{\pm i_m,\pm i_{m+1}}\right.\notag\\
&\left.\hspace{0mm}
-\xi_{\pm i_k^{\prime}}
\left(\prod_{s = 1}^{m - 1}D_{\pm i_s^{\prime}}\right)
r_{\pm i_m^{\prime},\pm i_{m+1}^{\prime}}\right\}.
\end{align}
}Every quantity on the right-hand side is either a prescribed composite retiming, a coupling coefficient determined by the phase-locking scheme, or a clock observable derived from measured sidebands. The unknown realization of $q$ has disappeared. Operationally, one parses the two signed operator sequences, selects the appropriate $r_{\pm i,\pm j}$ for every adjacent pair, applies the prefix preceding that pair, weights the result with $\xi_{\pm i_k}$, and finally differences the two branch accumulations.

For the complete data stream, a negative composite entry is implemented by the same measurable retiming, $\eta_{-i}\equiv-D_{-i}\eta_i$.  This extends the inverse-symbol definition from the clock projection to the GW and secondary-noise terms. Subtracting the template from the full phase-locking geometric-TDI data combination of Eq.~\eqref{interetaLASER} defines the calibrated observable
\par
{\small
\begin{align}\label{schem}
{\rm{TDI}}_{cancel - q}^{{\rm{geo}}} &={\rm TDI}_{\rm geo}-{\rm TDI}^q \notag\\
&=\sum\limits_{k = 1}^n \left\{
\begin{aligned}
&\left( {\prod\limits_{s = 1}^{k - 1} {{D_{ \pm {i_s}}}} } \right){\eta _{ \pm {i_k}}}
+ {\xi _{ \pm {i_k}}}\sum\limits_{m = 1}^{k - 1}
 {\left( {\prod\limits_{s = 1}^{m - 1} {{D_{ \pm {i_s}}}} } \right)
 {r_{ \pm {i_m}, \pm {i_{m + 1}}}}}\\
&- \left( {\prod\limits_{s = 1}^{k - 1} {{D_{ \pm {i_s^{\prime}}}}} } \right){\eta _{ \pm {i_k^{\prime}}}}
- {\xi _{ \pm {i_k^{\prime}}}}\sum\limits_{m = 1}^{k - 1}
 {\left( {\prod\limits_{s = 1}^{m - 1} {{D_{ \pm {i_s^{\prime}}}}} } \right)
 {r_{ \pm {i_m^{\prime}}, \pm {i_{m + 1}^{\prime}}}}}
\end{aligned}
\right\}.
\end{align}
}

Because Eq.~\eqref{schem} uses the complete $\eta_{\pm i}$ streams, the GW response and the remaining secondary noises are retained while the coefficient of the reference USO fluctuation cancels in the ideal model. The phase-locking-specific result is the algorithmic map from two sequences in the composite alphabet $\{a,b,c,d\}$ to a realizable clock template assembled from the corresponding sideband observables. It applies to a two-branch geometric-TDI solution only after the propagation-time and clock-coefficient closure checks have been satisfied. Equations~\eqref{kq}--\eqref{schem} assume that each branch is anchored with a positive first symbol, so that $q_{+i_1}=q$. If a branch begins with an inverse symbol while $q_{-i}=D_{-i}q$ is retained, its boundary contribution $q_{-i_1}-q=-D_{-i_1}r_{i_1}$ must be included before the template is applied. Finite phase-locking error, modulation noise, delay error, and imperfect coupling calibration set the residual in an implementation. Section~\ref{section6} evaluates this construction numerically.
\par
\begin{multicols}{2}

\section{Numerical simulations and clock-noise residuals}
\label{section6}

This section tests the clock-noise suppression algorithm developed in
this work by comparing time-domain simulations with the corresponding
frequency-domain models before and after clock calibration.  Two
16-link examples are considered in the phase-locking configuration,
$[X]^{16}_{1}$ and $[PE]^{16}_{1}$ \cite{tdi-geometric-2022}.  The former
contains only positive composite retimings, whereas the latter also contains
inverse retimings and therefore tests the complete signed-operator
construction of Eqs.~\eqref{newr}--\eqref{schem}.  For each geometric-TDI
solution, the case without phase-locking is generated from the
six one-way measurements.  The two architectures do not have identical raw
measurement combinations, so their residuals are compared with the transfer
model appropriate to each architecture.

\subsection{Numerical configuration}
\label{sec:simv2-configuration}

The carrier, sideband, test-mass, and reference measurements are generated
at $F_{\rm phy}=20~{\rm Hz}$ for two days using a time-dependent LISA orbit
and the corresponding six one-way light-travel times.  A 241-tap anti-alias
filter is applied before decimation to $F_{\rm dec}=4~{\rm Hz}$.  Samples
affected by the delay operations are discarded.  Time-dependent delays and
advances are evaluated with a 20-point, 19th-degree Lagrange fractional-delay
kernel.  The spectra are analyzed from $10^{-4}~{\rm Hz}$ to $1~{\rm Hz}$.
The simulated curves retain the time dependence of the delays.  The analytic
curves use arm lengths and beat-note couplings averaged over the same two-day
interval.

The optical carrier and nominal USO frequencies are
\begin{align}
 \nu_0=3\times10^{14}~{\rm Hz},\quad
 f_0=10~{\rm MHz},\notag\\
 g\equiv\frac{\nu_0}{f_0}=3\times10^7.
 \label{simv2-frequencies}
\end{align}
Thus, $g$ is the dimensionless ratio of the optical carrier frequency to
the nominal onboard-USO frequency.
The carrier--sideband modulation frequencies are $2.400~{\rm GHz}$ on
links $12$, $23$, and $31$, and $2.401~{\rm GHz}$ on links $13$, $21$,
and $32$.  These values correspond to the dimensionless modulation ratios
$m_{1,2,3}=(2.400~{\rm GHz})/f_0=240$ and
$m_{4,5,6}=(2.401~{\rm GHz})/f_0=240.1$ used when the sideband
measurements are formed.
The quantities $a_{ij}$ and $b_{ij}$ are the double-index beat-frequency
coefficients defined in Sec.~\ref{section2.1}.  We normalize them by the
nominal USO frequency as $\bar a_{ij}=a_{ij}/f_0$ and
$\bar b_{ij}=b_{ij}/f_0$.  The six optical-frequency offsets are
\begin{align}
 &(o_{12},o_{23},o_{31},\notag\\
 &\quad o_{13},o_{21},o_{32})\notag\\[-1mm]
 &\qquad=(8.1,9.2,10.3,1.4,-9.5,-11.6)~{\rm MHz}.
 \label{simv2-offsets}
\end{align}
They determine the zero-range-rate heterodyne coefficients and the
local-reference coefficients,
\begin{subequations}\label{simv2-beatcoefficients}
\begin{align}
 &(\bar a^{(0)}_{12},\bar a^{(0)}_{23},\bar a^{(0)}_{31},
   \bar a^{(0)}_{13},\bar a^{(0)}_{21},\bar a^{(0)}_{32})
 \notag\\[-1mm]
 &\hspace{5mm}=(-1.76,-2.08,-0.89,0.89,1.76,2.08),
 \label{simv2-aij-static}\\
 &(\bar b_{12},\bar b_{23},\bar b_{31},
   \bar b_{13},\bar b_{21},\bar b_{32})
 \notag\\[-1mm]
 &\hspace{5mm}=(-0.67,-1.87,-2.19,0.67,1.87,2.19).
 \label{simv2-bij-static}
\end{align}
\end{subequations}
For a time-dependent constellation, line-of-sight motion contributes to the
heterodyne coefficients.  Let $L_{ij}(t)$ be the optical path length from
spacecraft $j$ to spacecraft $i$ and let $d_{ij}=L_{ij}/c$ be the associated
light-travel time.  We define the dimensionless range rate and the normalized
heterodyne coefficient by
\begin{align}
 V_{ij}(t)&\equiv\frac{\dot L_{ij}(t)}{c}=\dot d_{ij}(t),\notag\\
 \bar a_{ij}(t)&=\bar a^{(0)}_{ij}
 -\frac{\nu_{ji}}{f_0}V_{ij}(t)\notag\\
 &\simeq\bar a^{(0)}_{ij}-gV_{ij}(t),
 \label{simv2-beatnormalization}
\end{align}
The superscript $(0)$ denotes the zero-range-rate value.  The last expression
uses $\nu_{ji}\simeq\nu_0$.  The orbital range-rate contribution is retained
in both architectures when their clock requirements are compared.

The input fractional-frequency amplitude spectral density (ASD) of each
physical USO is
\begin{align}
 S_y^{1/2}(f)=6.32\times10^{-14}
 \sqrt{\frac{1~{\rm Hz}}{f}}\ {\rm Hz}^{-1/2},
 \label{simv2-clockasd}
\end{align}
and the secondary-noise models are
\par
{\small
\begin{subequations}
\begin{align}
 S_x^{1/2}(f)={}&15
 \sqrt{1+\left(\frac{2~{\rm mHz}}{f}\right)^4}
 \ {\rm pm}\,{\rm Hz}^{-1/2},\\
 S_a^{1/2}(f)={}&3\times10^{-15}
 \sqrt{1+\left(\frac{0.4~{\rm mHz}}{f}\right)^2}
 \notag\\ &\qquad\times {\rm m}\,{\rm s}^{-2}{\rm Hz}^{-1/2}.
\end{align}
\label{simv2-secondarymodels}
\end{subequations}
}
The three USO fractional-frequency noises are generated independently and
have the same one-sided PSD $S_y(f)$, whose ASD is given by
Eq.~\eqref{simv2-clockasd}.  Their mutual cross-spectral densities therefore
vanish in the analytic model.

A monochromatic source is injected at $f_{\rm GW}=1~{\rm mHz}$ with
polarization angle $\psi=0$, detector-frame sky angles
$(\theta,\phi)=(60^\circ,180^\circ)$, and input strain ASD
$10^{-18}~{\rm Hz}^{-1/2}$.  The discrete sinusoidal amplitude is scaled as
$10^{-18}\sqrt{2F_{\rm phy}/N}$ for a record of $N$ physical-rate samples,
so the prescribed line amplitude is independent of the observation duration.
The numerical ASD is evaluated with a fast Fourier transform (FFT) as
\begin{align}
 \widehat{\mathcal A}_x(f_k)=
 \left|{\rm FFT}[w_nx_n]_k\right|
 \sqrt{\frac{2}{F_{\rm dec}N}},
 \label{simv2-asdestimator}
\end{align}
where $N$ is the number of samples in the usable record and
$w_n=\sin^8[\pi(n+1/2)/N]/(35/128)$.

\subsection{Clock-noise residuals with and without phase-locking}
\label{sec:simv2-general-transfer}

Hereafter, ``clock-noise residual'' denotes the clock contribution remaining
after the laser-noise-canceling TDI observable has been formed but before the
final sideband clock calibration is applied.  It is distinct from the much
smaller post-calibration residual generated by delay errors, noncommuting
delays, modulation noise, or readout noise.
For the phase-locking chain, the measured clock differences used in forming
the reduced streams have already referred the remote-clock contributions to
$q_1$.  Thus, ``before final clock calibration'' denotes the input to the last
sideband-template subtraction in the corresponding processing chain, rather
than identical raw carrier combinations in the two architectures.

We first consider the configuration without phase-locking.  With $D_{ij}$
propagating a quantity from spacecraft $j$ to spacecraft $i$, the clock parts
of the six effective one-way links are
\begin{subequations}\label{simv2-conventional-link-clocks}
\begin{align}
 \eta_{12}^{q}&=-f_0\!\left(\bar a_{12}q_1+
 \bar b_{21}D_{12}q_2\right),\notag\\
 \eta_{13}^{q}&=-f_0(\bar a_{13}-\bar b_{13})q_1,\\
 \eta_{23}^{q}&=-f_0\!\left(\bar a_{23}q_2+
 \bar b_{32}D_{23}q_3\right),\notag\\
 \eta_{21}^{q}&=-f_0(\bar a_{21}-\bar b_{21})q_2,\\
 \eta_{31}^{q}&=-f_0\!\left(\bar a_{31}q_3+
 \bar b_{13}D_{31}q_1\right),\notag\\
 \eta_{32}^{q}&=-f_0(\bar a_{32}-\bar b_{32})q_3.
\end{align}
\end{subequations}

The quantities $q_i$, $a_{ij}$, $b_{ij}$, and $D_{ij}$ retain the definitions
given in Sec.~\ref{section2.1}.  In particular, $q_i$ is the USO timing error
and $y_i=\dot q_i$ is its fractional-frequency fluctuation.  The frozen
transfer calculation retains the range-rate contribution in $a_{ij}$ and
uses the leading-order propagation of $D_{ij}\dot q_j$.  Terms of order
$V_{ij}y_j$ are neglected.

In the frozen-coefficient transfer model, terms proportional to
$\dot a_{ij}q_i$ and, below, $\dot\xi_\mu q_1$ are neglected.
For compact notation, the superscripts $\mathrm{PL}$ and $\mathrm{no\,PL}$
denote the configurations with and without phase-locking, respectively.
For an observable formed without phase-locking,
$\mathrm{TDI}=\sum_{i\ne j}P_{ij}\eta_{ij}$, differentiation of the phase
observable and normalization by $\nu_0$ give
\begin{align}
 \widetilde{\mathrm{TDI}}_{y,q}^{\rm no\,PL}(f)
 =-\frac{f_0}{\nu_0}\sum_{i=1}^{3}
 H_i^{\rm no\,PL}(f)\widetilde y_i(f),
 \label{simv2-conventional-clock-residual}
\end{align}
where
\begin{subequations}
\begin{align}
 H_{1}^{\rm no\,PL}={}&
 \bar a_{12}\widetilde P_{12}
 +(\bar a_{13}-\bar b_{13})\widetilde P_{13}
 +\bar b_{13}\widetilde P_{31}\widetilde D_{31},\\
 H_{2}^{\rm no\,PL}={}&
 \bar a_{23}\widetilde P_{23}
 +(\bar a_{21}-\bar b_{21})\widetilde P_{21}
 +\bar b_{21}\widetilde P_{12}\widetilde D_{12},\\
 H_{3}^{\rm no\,PL}={}&
 \bar a_{31}\widetilde P_{31}
 +(\bar a_{32}-\bar b_{32})\widetilde P_{32}
 +\bar b_{32}\widetilde P_{23}\widetilde D_{23}.
\end{align}
\label{simv2-conventional-transfers}
\end{subequations}
Equations~\eqref{simv2-conventional-clock-residual} and
\eqref{simv2-conventional-transfers} have the same coefficient pattern as
Eq.~(18) of Ref.~\cite{tdi-clock-pan} after its single-link indices are
translated to the present convention, in which $ij$ denotes reception at
spacecraft $i$ of light transmitted by spacecraft $j$.  The time-domain operator form does not
require equal arms.  Scalar Fourier transfers require frozen delays, which
may still be unequal.  Only in the frozen equal-arm specialization do
$\widetilde D_{12}=\widetilde D_{23}=\widetilde D_{31}=z=e^{iu}$, with
$u=2\pi fL/c$.

For the phase-locking configuration, the four effective streams satisfy
$(\eta_1,\eta_2,\eta_{1'},\eta_{3'})=(\eta_b,\eta_c,\eta_a,\eta_d)$.
The corresponding clock couplings introduced in Sec.~\ref{section5.1} are
therefore relabeled as
$\xi_1\equiv\xi_b$, $\xi_2\equiv\xi_c$,
$\xi_{1'}\equiv\xi_a$, and $\xi_{3'}\equiv\xi_d$ when the streams carry
numerical rather than composite labels.  These equalities are only a change
of labels.  In terms of the double-index coefficients of
Sec.~\ref{section2.1},
\begin{align}
 (\xi_1,\xi_2,\xi_{1'},\xi_{3'})={}&\notag\\
 & \bigl(a_{12}+a_{21},\,\notag\\
 &
 a_{23}-b_{31}+a_{31}-b_{12}\notag\\ &-a_{21}-b_{23},\notag\\
 &a_{13}+a_{31},\,\notag\\
 &
 a_{32}+b_{31}-a_{31}+a_{21}\notag\\ &+b_{23}+b_{12}\bigr).
 \label{simv2-xi-explicit}
\end{align}

Because $a_{ij}$ and $b_{ij}$ have units of hertz, every $\xi_\mu$ in
Eq.~\eqref{simv2-xi-explicit} also has units of hertz.  Only the normalized
coefficient
\begin{align}
 \bar\xi_\mu(t)\equiv\frac{\xi_\mu(t)}{f_0},
 \qquad \mu\in\{1,2,1',3'\}.
 \label{simv2-xinormalization}
\end{align}
is dimensionless.  For arbitrary Fourier-domain weights
$\widetilde P_\mu$, the phase-locking clock residual is
\begin{subequations}
\begin{align}
 \widetilde{\mathrm{TDI}}_{y,q}^{\rm PL}(f)={}&
 -\frac{f_0}{\nu_0}\bar H_q(f)\widetilde y_1(f),\\
 \bar H_q(f)={}&
 \widetilde P_1\bar\xi_1+
 \widetilde P_2\bar\xi_2+
 \widetilde P_{1'}\bar\xi_{1'}+
 \widetilde P_{3'}\bar\xi_{3'}.
\end{align}
\label{simv2-generalHq}
\end{subequations}
Referring the remote-clock differences to spacecraft 1 through the sideband
measurements reduces the clock sector to one transfer column in the ideal
phase-locking representation.

For the general case without phase-locking, introduce the one-sided cross-spectral
matrix $\mathbf S_y(f)=[S_{y,ij}(f)]$ and the row vector
$\mathbf H_{\rm no\,PL}=(H_1^{\rm no\,PL},H_2^{\rm no\,PL},H_3^{\rm no\,PL})$.
We use
$\langle\widetilde y_i(f)\widetilde y_j^*(f')\rangle
=\tfrac12\delta(f-f')S_{y,ij}(f)$.
The clock-noise PSDs before final clock calibration are
\begin{subequations}\label{simv2-generalclock-spectrum}
\begin{align}
 S_{\mathrm{TDI},q}^{\rm no\,PL}(f)={}&
 \left(\frac{f_0}{\nu_0}\right)^2
 \mathbf H_{\rm no\,PL}\mathbf S_y\mathbf H_{\rm no\,PL}^{\dagger},\\
 S_{\mathrm{TDI},q}^{\rm PL}(f)={}&
 \left(\frac{f_0}{\nu_0}\right)^2
 |\bar H_q(f)|^2S_{y,11}(f).
\end{align}
\end{subequations}
The first line contains the mutual-USO cross terms whenever
$S_{y,ij}\ne0$.  In the second line, the four stream contributions must be
added coherently before the modulus is taken because each multiplies the same
reference process $y_1$.  For real frozen couplings,
\begin{align}
 |\bar H_q|^2={}&
 \sum_\mu\bar\xi_\mu^2|\widetilde P_\mu|^2
 +2\sum_{\mu<\nu}\bar\xi_\mu\bar\xi_\nu
 {\rm Re}\!\left(\widetilde P_\mu\widetilde P_\nu^*\right).
 \label{simv2-coherentpower}
\end{align}

For mutually independent USOs with a common spectrum $S_y$, define
\begin{align}
 K_{\rm no\,PL}(f)\equiv
 \left(\sum_{i=1}^{3}|H_i^{\rm no\,PL}(f)|^2\right)^{1/2}.
 \label{simv2-Kconventional}
\end{align}
Equation~\eqref{simv2-generalclock-spectrum} then reduces to
\begin{subequations}
\begin{align}
 \left[S_{\mathrm{TDI},q}^{\rm no\,PL}(f)\right]^{1/2}={}&
 \frac{f_0}{\nu_0}K_{\rm no\,PL}(f)S_y^{1/2}(f),\\
 \left[S_{\mathrm{TDI},q}^{\rm PL}(f)\right]^{1/2}={}&
 \frac{f_0}{\nu_0}|\bar H_q(f)|S_y^{1/2}(f).
\end{align}
\label{simv2-identical-clock-asds}
\end{subequations}
The first line of Eq.~\eqref{simv2-generalclock-spectrum} then recovers
Eq.~(19) of Ref.~\cite{tdi-clock-pan} after conversion from clock phase to
fractional frequency.
For possibly different secondary-noise targets in the two architectures, the
corresponding maximum input-clock ASDs and their ratio are
\begin{subequations}\label{simv2-clock-requirements}
\begin{align}
 \left[S_{y,{\rm req}}^{\rm PL}(f)\right]^{1/2}={}&
 \frac{\nu_0}{f_0}
 \frac{\left[S_{{\rm TDI},{\rm sec}}^{\rm PL}(f)\right]^{1/2}}{|\bar H_q(f)|},\\
 \left[S_{y,{\rm req}}^{\rm no\,PL}(f)\right]^{1/2}={}&
 \frac{\nu_0}{f_0}
 \frac{\left[S_{{\rm TDI},{\rm sec}}^{\rm no\,PL}(f)\right]^{1/2}}{K_{\rm no\,PL}(f)},\\
 \mathcal G(f)\equiv
 \left[\frac{S_{y,{\rm req}}^{\rm PL}}
      {S_{y,{\rm req}}^{\rm no\,PL}}\right]^{1/2}
 ={}&\left[\frac{S_{{\rm TDI},{\rm sec}}^{\rm PL}}
 {S_{{\rm TDI},{\rm sec}}^{\rm no\,PL}}\right]^{1/2}
 \frac{K_{\rm no\,PL}(f)}{|\bar H_q(f)|}.
 \label{simv2-relaxation-factor}
\end{align}
\end{subequations}
Here $\mathcal G$ is the ratio of the maximum common input ASD assigned to the
three physical onboard USOs in the two processing chains defined above.  In
Eq.~\eqref{simv2-generalHq}, the measured clock-difference observables refer
the contributions of the spacecraft-2 and spacecraft-3 USOs to the
spacecraft-1 clock.  The phase-locking clock sector is therefore represented
by a single transfer function multiplying $y_1$.  This algebraic reduction
does not imply that the three physical USOs are synchronized or correlated.
Thus, as defined in Eq.~\eqref{simv2-clock-requirements}, $\mathcal G>1$
means that the phase-locking processing chain admits a larger maximum input
clock ASD for its specified secondary-noise target.  Its equivalent
clock-residual interpretation for the paired equal-arm comparison is given
below.

For each paired realization in the catalog comparison, expanding its four
phase-locking streams into six one-way links gives the same GW, test-mass, and
carrier optical-path noise transfer functions in the ideal equal-arm model.
Noise associated with the auxiliary sideband and ranging operations is not
part of this secondary-noise target.  The clock transfer remains different
because the phase-locking reduction refers the remote-clock contributions to
the spacecraft-1 clock process.  Under these assumptions,
Eq.~\eqref{simv2-relaxation-factor} reduces to
$\mathcal G=K_{\rm no\,PL}/|\bar H_q|$.  For the same common input clock
spectrum, Eqs.~\eqref{simv2-identical-clock-asds} and
\eqref{simv2-relaxation-factor} then give
$\mathcal G=[S_{\mathrm{TDI},q}^{\rm no\,PL}/
S_{\mathrm{TDI},q}^{\rm PL}]^{1/2}$.  Thus, in the paired equal-arm
comparison used below, $\mathcal G$ directly measures the reduction of the
pre-calibration clock-noise residual ASD produced by phase-locking.  The
corresponding PSD reduction factor is $\mathcal G^2$.

In the time-domain calculations, the residual without phase-locking in
Eq.~\eqref{simv2-conventional-clock-residual} is calibrated with the standard
sideband construction of Ref.~\cite{tdi-clock-pan}, whereas the
phase-locking residual is calibrated with Eq.~\eqref{schem}.
In the time-domain simulations, the clock-noise residual after calibration
lies below the secondary-noise floor formed by test-mass acceleration noise
and optical-path noise for both architectures over the analyzed frequency
band.  The simulated curves retain the dynamic delays and fractional-delay
interpolation.

\subsection{Range-rate dependence of the effective couplings}
\label{sec:simv2-doppler-couplings}

Here $L_{ij}(t)$ is the optical path length from spacecraft $j$ to
spacecraft $i$, and $V_{ij}$, defined in
Eq.~\eqref{simv2-beatnormalization}, is its dimensionless range rate.
Positive $V_{ij}$ denotes an increasing optical path.  For the frequency
plan in Eq.~\eqref{simv2-offsets}, the constant parts formed from
$\bar a_{ij}$ and $\bar b_{ij}$ cancel in each effective coupling.  The
remaining terms are
\begin{subequations}\label{simv2-Dopplercouplings}
\begin{equation}
\begin{aligned}
 \bar\xi_1={}&\bar a_{12}+\bar a_{21}\\
 \simeq{}&-g(V_{12}+V_{21}).
\end{aligned}
\end{equation}
\begin{equation}
\begin{aligned}
 \bar\xi_2={}&\bar a_{23}-\bar b_{31}+\bar a_{31}
 -\bar b_{12}-\bar a_{21}-\bar b_{23}\\
 \simeq{}&g(-V_{23}-V_{31}+V_{21}).
\end{aligned}
\end{equation}
\begin{equation}
\begin{aligned}
 \bar\xi_{1'}={}&\bar a_{13}+\bar a_{31}\\
 \simeq{}&-g(V_{13}+V_{31}).
\end{aligned}
\end{equation}
\begin{equation}
\begin{aligned}
 \bar\xi_{3'}={}&\bar a_{32}+\bar b_{31}-\bar a_{31}
 +\bar a_{21}+\bar b_{23}+\bar b_{12}\\
 \simeq{}&g(-V_{32}+V_{31}-V_{21}).
\end{aligned}
\end{equation}
\end{subequations}
Equation~\eqref{simv2-Dopplercouplings} explains why the effective
coefficients can be of order unity even though $|V_{ij}|\ll1$, because the
range rates are multiplied by $g=3\times10^7$.  It also shows that a strictly
static constellation gives $\bar\xi_\mu=0$ for this ideal frequency plan.
The frozen analytic calculation does not impose that static limit.  It
retains the orbital means of the range-rate combinations while freezing their
time dependence over a Fourier transform.

\end{multicols}
For the two-day orbit, each range-rate sample is averaged over the usable
record according to
$\langle V_{ij}\rangle=N^{-1}\sum_{n=0}^{N-1}V_{ij}(t_n)$ and then
substituted into Eq.~\eqref{simv2-Dopplercouplings}.  The resulting averaged
range-rate combinations and effective coefficients are
\begin{align}
 \begin{array}{c|c|c}
 \text{stream}&\text{averaged range-rate combination}&
 \langle\bar\xi_\mu\rangle\\ \hline
 \eta_1&\langle V_{12}+V_{21}\rangle=-2.99442\times10^{-8}
 &0.8983261\\
 \eta_2&\langle -V_{23}-V_{31}+V_{21}\rangle=2.02046\times10^{-8}
 &0.6061394\\
 \eta_{1'}&\langle V_{13}+V_{31}\rangle=-4.20811\times10^{-8}
 &1.2624339\\
 \eta_{3'}&\langle -V_{32}+V_{31}-V_{21}\rangle=8.06772\times10^{-9}
 &0.2420316
 \end{array}.
 \label{simv2-mean-coupling-table}
\end{align}

The table is the numerical evaluation of
Eq.~\eqref{simv2-Dopplercouplings} and converts the two-day mean range rates
into the four dimensionless couplings multiplying the common reference-clock
process.  The right column is the time average of the four coefficients
constructed in Eq.~\eqref{simv2-Dopplercouplings}.  Applying the same two-day
mean range rates to the six coefficients used by the transfer without
phase-locking gives, in the ordering of
Eq.~\eqref{simv2-beatcoefficients},
\begin{multicols}{2}
\begin{align}
 &(\langle\bar a_{12}\rangle,\langle\bar a_{23}\rangle,
  \langle\bar a_{31}\rangle,\notag\\
 &\quad\langle\bar a_{13}\rangle,
  \langle\bar a_{21}\rangle,\langle\bar a_{32}\rangle)\notag\\[-1mm]
 &\qquad=(-1.310827,-1.655907,-0.258781,\notag\\
 &\hspace{14mm}1.521219,2.209173,2.504093).
 \label{simv2-mean-aij}
\end{align}
These are the complete frozen coefficients, including the range-rate
contribution.  They are not the zero-range-rate values of
Eq.~\eqref{simv2-beatcoefficients}.  The combinations that enter the two
phase-locking examples are
\begin{align}
 \Delta_{11'}&\equiv
 \langle\bar\xi_1\rangle-\langle\bar\xi_{1'}\rangle
 =-0.3641078,\notag\\
 \Delta_{23'}&\equiv
 \langle\bar\xi_2\rangle-\langle\bar\xi_{3'}\rangle
 =0.3641078.
 \label{simv2-delta-couplings}
\end{align}
Their nearly equal magnitudes and opposite signs are a consequence of the
chosen orbit average and frequency plan, rather than a general identity.

Using the two-day averaged coefficients in
Eqs.~\eqref{simv2-mean-coupling-table} and \eqref{simv2-mean-aij}, we evaluated
Eq.~\eqref{simv2-relaxation-factor} for the 45 distinct second-generation
geometric-TDI combinations with up to 16 links considered
here~\cite{geome-tdi-2023,phase-locking-tdi-2024}.  The calculation uses the
frozen equal-arm transfer model with $L=2.5\times10^9~\mathrm{m}$.
Figure~\ref{fig:all45-gain} summarizes the resulting frequency-dependent
clock-noise residual reduction factors.

\end{multicols}
\begin{inlinefigure}
\centering
\includegraphics[width=0.94\textwidth]{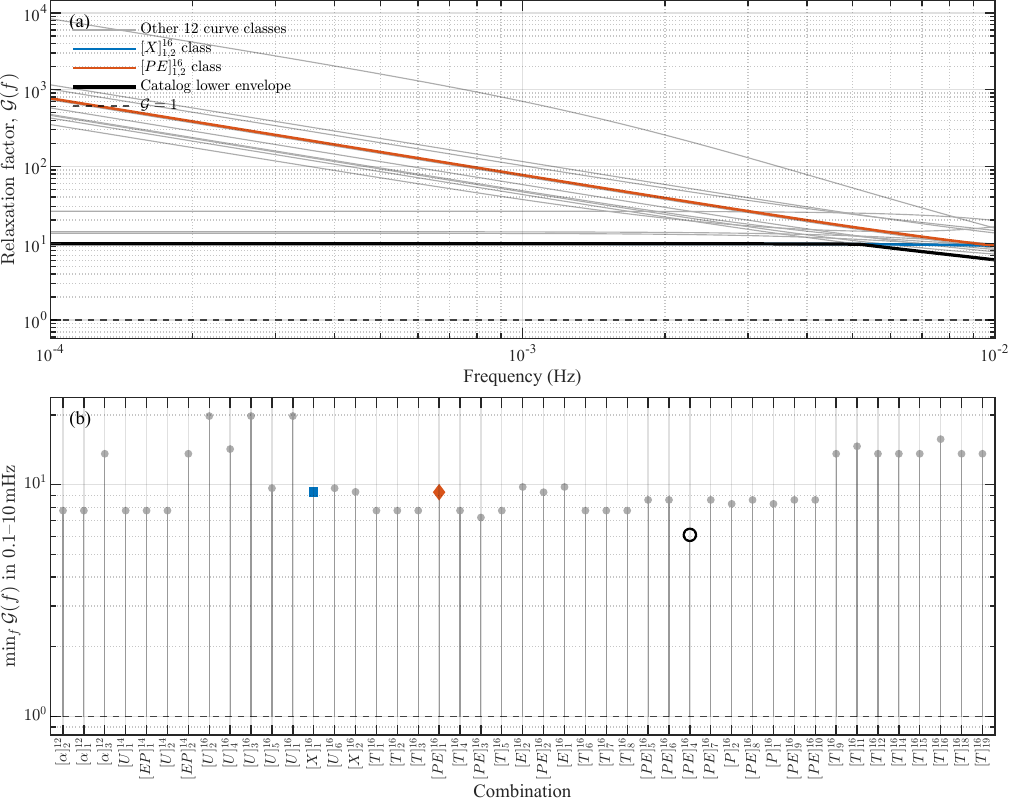}
\caption{\label{fig:all45-gain}
Clock-noise residual ASD reduction factor $\mathcal G$ for the 45 distinct
second-generation geometric-TDI combinations with up to 16 links considered
here.  Panel (a) shows one representative from each of the 14 numerical curve
classes over $0.1$--$10~\mathrm{mHz}$.  The 12 gray curves represent the
remaining classes.  The classes $\{[X]^{16}_{1},[X]^{16}_{2}\}$ and
$\{[PE]^{16}_{1},[PE]^{16}_{2}\}$ are highlighted.  The black curve is the lower envelope over
the catalog, and the horizontal dashed line marks $\mathcal G=1$.  Panel (b)
shows the minimum of $\mathcal G(f)$ in this frequency interval for all 45
combinations, which are identified on the horizontal axis.  Repeated values
correspond to entries in the same numerical
curve class.  The open circle marks the catalog-wide minimum
$\mathcal G=6.080$ at $10~\mathrm{mHz}$ for $[PE]^{16}_{4}$.  All 45 minima
exceed unity.}
\end{inlinefigure}
\begin{multicols}{2}

Under these assumptions, the 45 catalog entries fall into 14 classes whose
$\mathcal G(f)$ curves coincide to numerical precision over
$0.1$--$10~\mathrm{mHz}$.  Using the maximum pointwise relative difference
over $0.1$--$10~\mathrm{mHz}$ as the curve-distance measure, the largest
within-class distance is $2.1\times10^{-11}$, whereas the smallest distance
between two distinct classes is $7.4\times10^{-2}$.  The numerical classes are
therefore well separated under the adopted frozen equal-arm model.  The first
class contains
$[\alpha]_{1}^{12}$, $[\alpha]_{2}^{12}$, $[U]_{1}^{14}$,
$[U]_{2}^{14}$, $[EP]_{1}^{14}$, and $[T]_{k}^{16}$ for
$1\leq k\leq8$.  The second class contains $[\alpha]_{3}^{12}$,
$[EP]_{2}^{14}$, and $[T]_{k}^{16}$ for
$k=9,12,14,15,18,19$.  The other multi-entry classes are
$\{[U]_{1}^{16},[U]_{2}^{16},[U]_{3}^{16}\}$,
$\{[U]_{5}^{16},[U]_{6}^{16}\}$,
$\{[X]_{1}^{16},[X]_{2}^{16}\}$,
$\{[PE]_{1}^{16},[PE]_{2}^{16}\}$,
$\{[E]_{1}^{16},[E]_{2}^{16}\}$,
$\{[PE]_{5}^{16},[PE]_{6}^{16},[PE]_{7}^{16},
[PE]_{8}^{16},[PE]_{9}^{16},[PE]_{10}^{16}\}$, and
$\{[P]_{1}^{16},[P]_{2}^{16}\}$.  The five single-entry classes are
$[U]_{4}^{16}$, $[PE]_{3}^{16}$, $[PE]_{4}^{16}$,
$[T]_{11}^{16}$, and $[T]_{16}^{16}$.  This classification concerns the
clock-noise residual reduction factors in the adopted frozen equal-arm model.

All 45 catalog entries remain above $\mathcal G=1$ on the evaluated
$0.1$--$10~\mathrm{mHz}$ grid.  The catalog-wide lower envelope reaches
$\mathcal G=6.080$ at $10~\mathrm{mHz}$ for $[PE]^{16}_{4}$.  Thus, under the
assumptions of this comparison and for the same input clock spectrum, every
phase-locking realization reduces the clock-noise residual ASD by at least a
factor of $6.080$ in the millihertz band considered here.  The following two
examples provide detailed time-domain tests of the analytical transfer
functions.

\subsection{The \texorpdfstring{$[X]^{16}_{1}$}{[X]16-1} example}
\label{sec:simv2-X}

Using the composite-operator mapping
\begin{align}
 \eta_a\equiv\eta_{1'},\qquad
 \eta_b\equiv\eta_1,\qquad
 \eta_c\equiv\eta_2,\qquad
 \eta_d\equiv\eta_{3'},
 \label{simv2-composite-stream-map}
\end{align}
the two phase-locking branches of $[X]^{16}_{1}$ are
\begin{subequations}
\begin{align}
 \mathcal B_X^{(1)}={}&
 \eta_b+D_b\eta_a+D_bD_a\eta_a+D_bD_aD_a\eta_b,\\
 \mathcal B_X^{(2)}={}&
 \eta_a+D_a\eta_b+D_aD_b\eta_b+D_aD_bD_b\eta_a,
\end{align}
\label{simv2-X-branches}
\end{subequations}
and
\begin{align}
 [X]^{16}_{1}=\mathcal B_X^{(1)}-\mathcal B_X^{(2)}.
 \label{simv2-X-flow}
\end{align}
This is the explicit eight-term signed composite-stream representation.

In the equal-arm approximation, $D_a=D_b=z^2$ and the four stream
polynomials are
\begin{align}
 \widetilde P_1^X={}(1-z^2)(1-z^4),\notag\\
 \widetilde P_{1'}^X={}&-(1-z^2)(1-z^4),\notag\\
 \widetilde P_2^X={}&0,\quad
 \widetilde P_{3'}^X={}&0.
 \label{simv2-X-polynomials}
\end{align}
Substitution into the general expression
Eq.~\eqref{simv2-generalHq}, rather than application of an additional TDI
identity, gives
\begin{subequations}
\begin{align}
 \bar H_{q,X}^{\rm PL}={}&
 (1-z^2)(1-z^4)\Delta_{11'},\\
 |\bar H_{q,X}^{\rm PL}|^2={}&
 16\sin^2u\,\sin^2(2u)\Delta_{11'}^2.
\end{align}
\label{simv2-X-clock-transfer}
\end{subequations}
The clock-noise power response is $O(u^4)$ at low frequency.
The realization without phase-locking associated with
$[X]^{16}_{1}$ has the one-way polynomials
\begin{align}
 &(\widetilde P_{12},\widetilde P_{23},\widetilde P_{31},\notag\\ &
  \widetilde P_{13},\widetilde P_{21},\widetilde P_{32})_X
\notag\\
 ={}&(A_X,0,-zA_X,-A_X,zA_X,0),\notag\\
 A_X\equiv{}&(1-z^2)(1-z^4),
 \label{simv2-X-conventional-polynomials}
\end{align}

which are inserted into Eq.~\eqref{simv2-conventional-transfers} with the
complete coefficients of Eq.~\eqref{simv2-mean-aij}.  In the ideal equal-arm
model, these polynomials are also obtained by expanding the four composite
phase-locking streams.  The paired realizations therefore have the same GW
and secondary-noise transfers.  Their clock-noise transfers differ according
to Eqs.~\eqref{simv2-conventional-transfers} and
\eqref{simv2-X-clock-transfer}.

Over the full $10^{-4}$--$1~\mathrm{Hz}$ interval used for this example,
$\mathcal G_X(f)>1$, with the conservative value
\begin{align}
 \min_f\mathcal G_X(f)=7.356.
 \label{simv2-X-results}
\end{align}
Thus, for the same input clock ASD, the phase-locking realization reduces the
clock-noise residual ASD by at least a factor of $7.356$.  Equivalently, if the
secondary-noise floor is used as the comparison target, the maximum admissible
input clock ASD is larger by the same factor.
Figures~\ref{fig:simv2-X-output}
and~\ref{fig:simv2-X-strain} show the corresponding output-noise and
equivalent-strain spectra.

\begin{inlinefigure}
\centering
\includegraphics[width=\columnwidth,trim=45 15 75 40,clip]{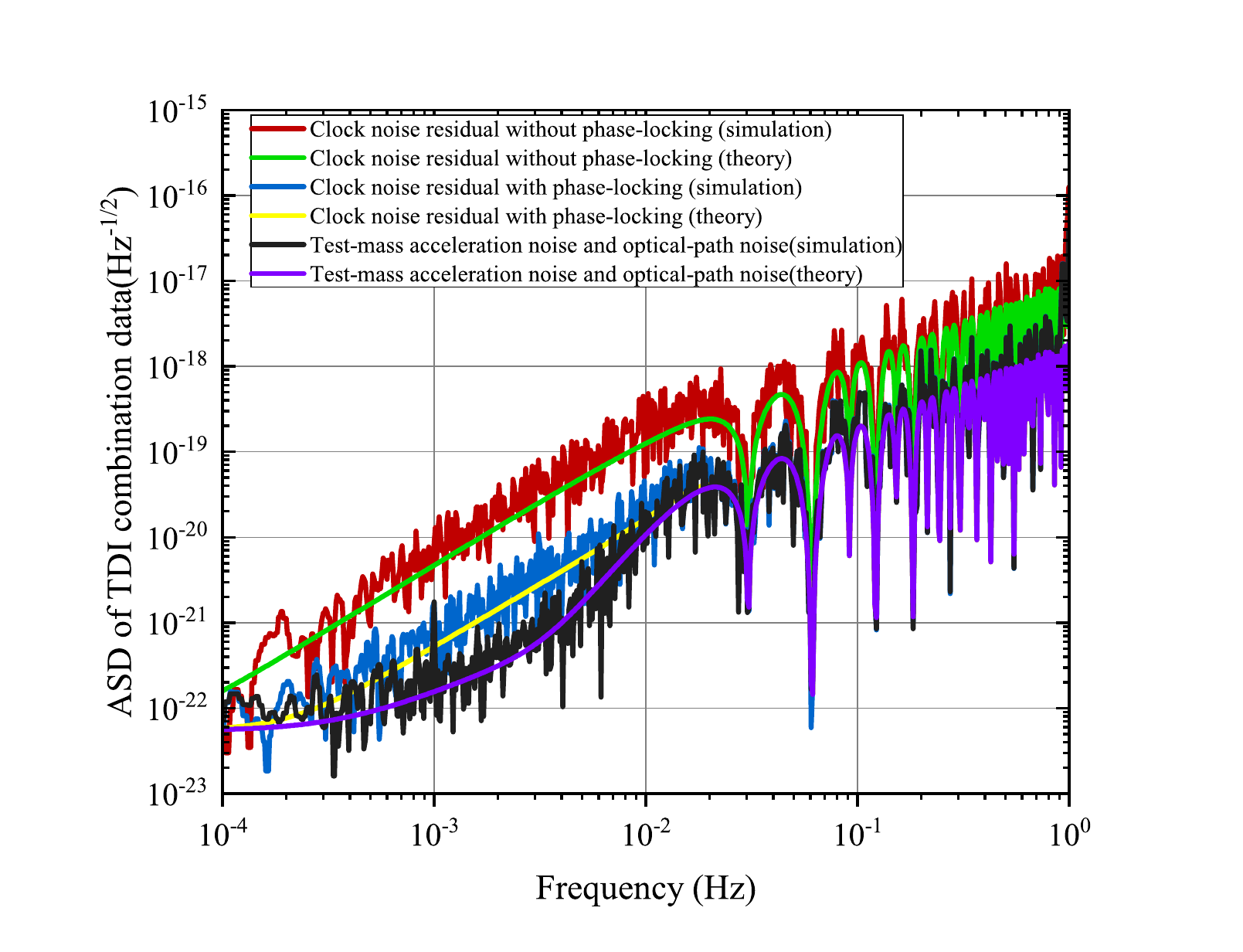}
\caption{\label{fig:simv2-X-output}
Output ASDs associated with $[X]^{16}_{1}$.  The simulated and analytic
clock-noise residuals are shown for the realizations with and without
phase-locking.  The simulated and analytic secondary-noise spectra contain
test-mass acceleration noise and optical-path noise.}
\end{inlinefigure}

\begin{inlinefigure}
\centering
\includegraphics[width=\columnwidth,trim=45 15 75 40,clip]{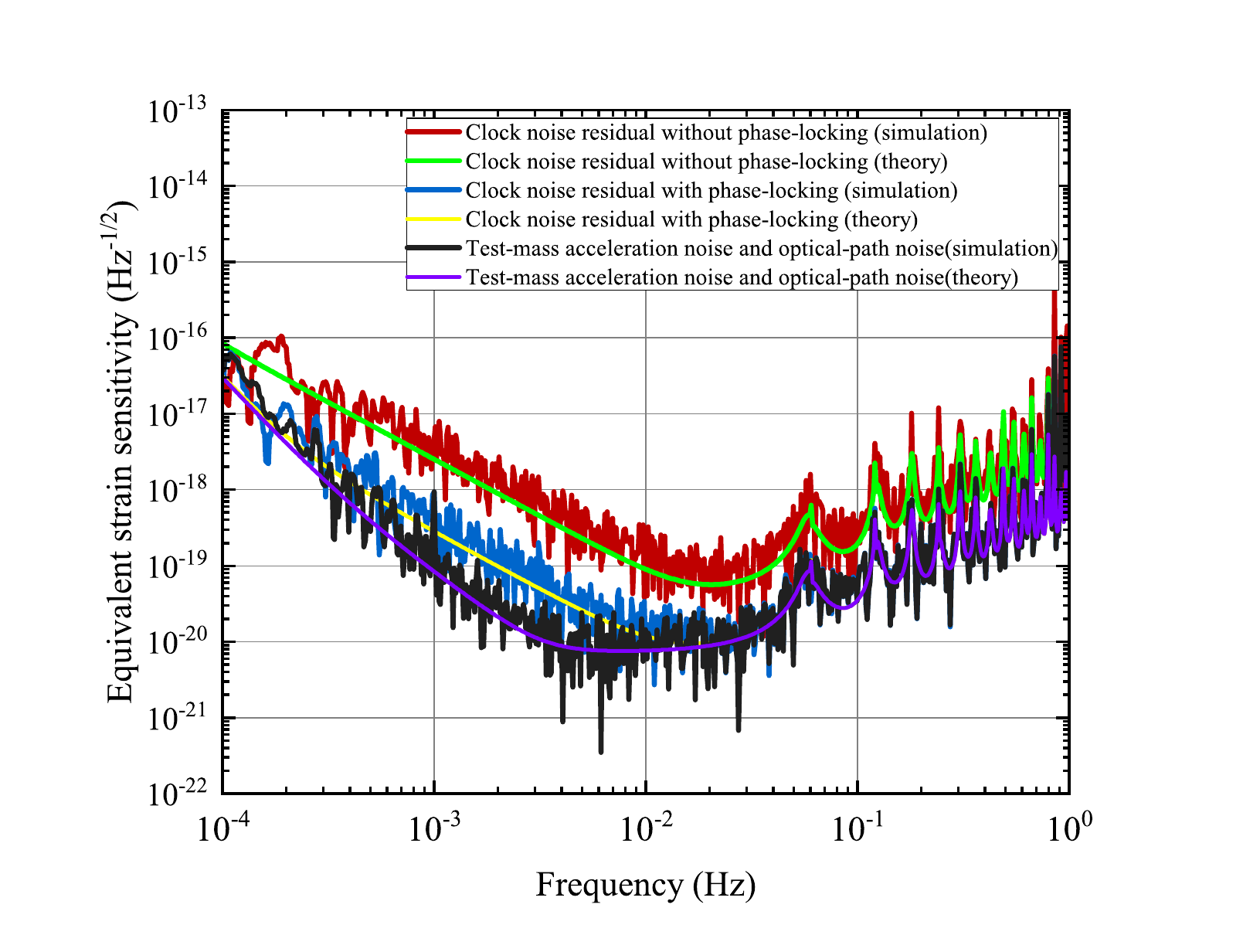}
\caption{\label{fig:simv2-X-strain}
Equivalent-strain ASDs associated with $[X]^{16}_{1}$, obtained by dividing
each output ASD by the absolute GW response of the corresponding realization.
The six curves have the same definitions as in
Fig.~\ref{fig:simv2-X-output}.}
\end{inlinefigure}

\subsection{The \texorpdfstring{$[PE]^{16}_{1}$}{[PE]16-1} example}
\label{sec:simv2-PE}

With $\eta_{-i}\equiv-D_{-i}\eta_i$, the two phase-locking branches of
$[PE]^{16}_{1}$ are
\par
{\small
\begin{subequations}
\begin{align}
 \mathcal B_{PE}^{(1)}={}&
 \eta_b+D_b\eta_c+D_bD_c\eta_d
 +D_bD_cD_d\eta_{-c}\notag\\
 &+D_bD_cD_dD_{-c}\eta_{-d}
 +D_bD_cD_dD_{-c}D_{-d}\eta_c,\\
 \mathcal B_{PE}^{(2)}={}&
 \eta_a+D_a\eta_d+D_aD_d\eta_c
 +D_aD_dD_c\eta_{-a}\notag\\
 &+D_aD_dD_cD_{-a}\eta_b
 +D_aD_dD_cD_{-a}D_b\eta_{-d},
\end{align}
\label{simv2-PE-branches}
\end{subequations}
}
and
\begin{align}
 [PE]^{16}_{1}=\mathcal B_{PE}^{(1)}-\mathcal B_{PE}^{(2)}.
 \label{simv2-PE-flow}
\end{align}
The expression contains twelve signed composite-stream entries and makes
the inverse retimings explicit.

For equal arms, $D_a=D_b=z^2$ and $D_c=D_d=z$.  The stream polynomials are
\begin{align}
 \widetilde P_1^{PE}={}&1-z^2,&
 \widetilde P_{1'}^{PE}={}&-(1-z^2),\notag\\
 \widetilde P_2^{PE}={}&2z^2(1-z),&
 \widetilde P_{3'}^{PE}={}&-2z^2(1-z).
 \label{simv2-PE-polynomials}
\end{align}
The corresponding specialization of Eq.~\eqref{simv2-generalHq} is
\begin{subequations}
\begin{align}
 \bar H_{q,PE}^{\rm PL}={}&
 (1-z^2)\Delta_{11'}+2z^2(1-z)\Delta_{23'},\\
 |\bar H_{q,PE}^{\rm PL}|^2={}&
 16\sin^2\!\left(\frac{u}{2}\right)
 \left[\cos^2\!\left(\frac{u}{2}\right)\Delta_{11'}^2
 +\Delta_{23'}^2\right.\notag\\
 &\left.\hspace{2mm}
 +(\cos u+\cos2u)\Delta_{11'}\Delta_{23'}\right].
\end{align}
\label{simv2-PE-clock-transfer}
\end{subequations}
For arbitrary couplings, the power response in
Eq.~\eqref{simv2-PE-clock-transfer} is $O(u^2)$.  The near relation
$\Delta_{23'}\simeq-\Delta_{11'}$ in
Eq.~\eqref{simv2-delta-couplings} produces the additional factorization
\begin{align}
 \bar H_{q,PE}^{\rm PL}\simeq
 \Delta_{11'}(1-z)^2(1+2z),
 \label{simv2-PE-additional-cancellation}
\end{align}
and an $O(u^4)$ power response for this particular frequency plan.  This
additional cancellation is not part of the general result
Eq.~\eqref{simv2-generalHq}.
The realization without phase-locking associated with
$[PE]^{16}_{1}$ has the one-way polynomials
\begin{equation}\label{simv2-PE-conventional-polynomials}
\begin{aligned}
 (\widetilde P_{12},\widetilde P_{23},\widetilde P_{31})_{PE}
 &=(A_{PE},C_{PE},zA_{PE}),\\
 (\widetilde P_{13},\widetilde P_{21},\widetilde P_{32})_{PE}
 &=(-A_{PE},-zA_{PE},-C_{PE}),\\
 A_{PE}&=1-z^2,\qquad C_{PE}=2z(1-z).
\end{aligned}
\end{equation}
The same polynomials follow from expansion of the four phase-locking streams
in the ideal equal-arm model.  The paired realizations therefore have the
same GW and secondary-noise transfers, whereas their clock-noise transfers
remain those of Eqs.~\eqref{simv2-conventional-transfers} and
\eqref{simv2-PE-clock-transfer}.

Over the full $10^{-4}$--$1~\mathrm{Hz}$ interval used for this example,
$\mathcal G_{PE}(f)>1$, with the conservative value
\begin{align}
 \min_f\mathcal G_{PE}(f)=5.351.
 \label{simv2-PE-results}
\end{align}
\begin{inlinefigure}
\centering
\includegraphics[width=\columnwidth,trim=45 15 75 40,clip]{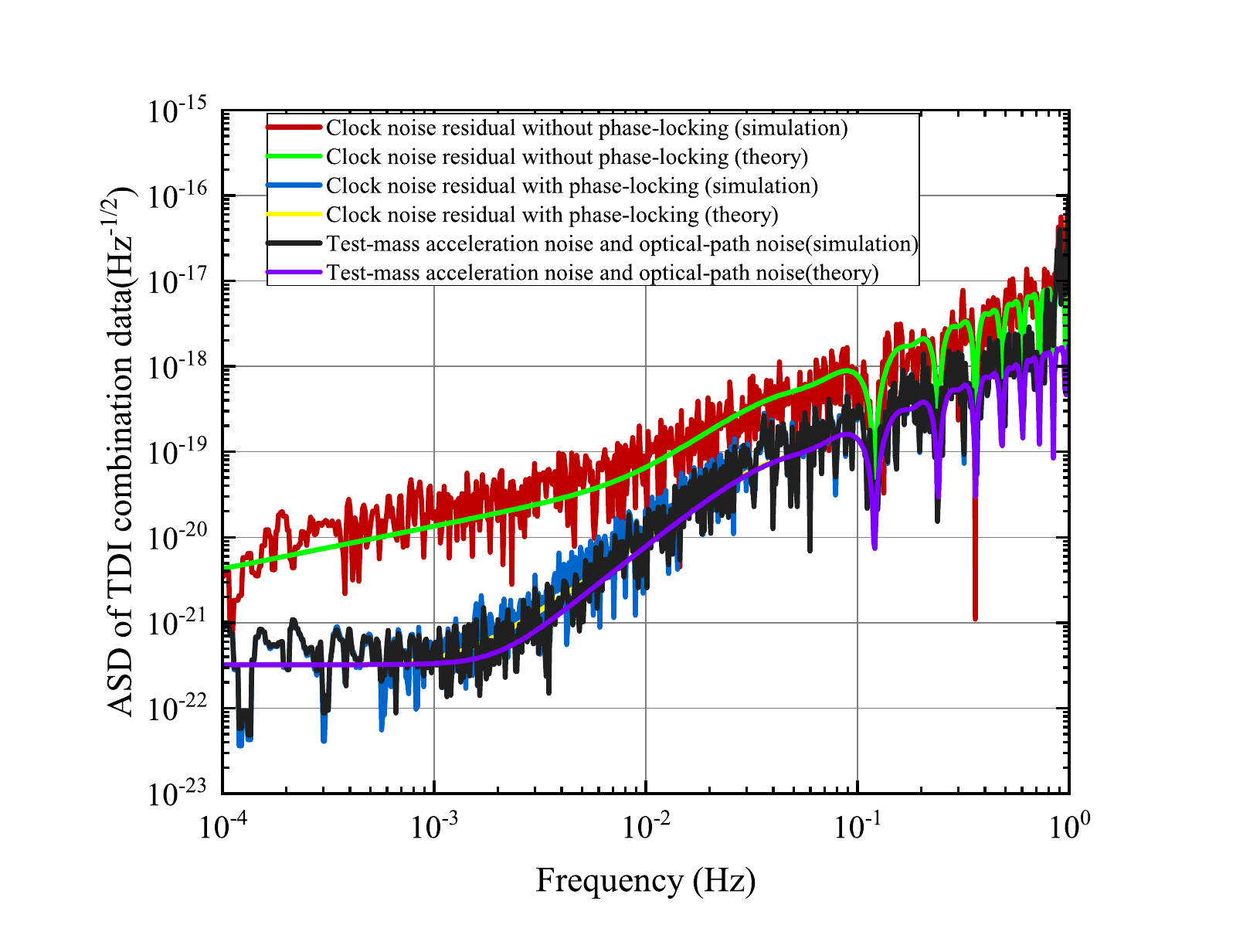}
\caption{\label{fig:simv2-PE-output}
Output ASDs for the phase-locking realization and the realization without
phase-locking associated with the $[PE]^{16}_{1}$ observable.  The simulated
and analytic clock-noise residuals are shown for both realizations.  The
simulated and analytic secondary-noise spectra contain test-mass acceleration
noise and optical-path noise.}
\end{inlinefigure}

Thus, for the same input clock ASD, the phase-locking realization reduces the
clock-noise residual ASD by at least a factor of $5.351$.  Equivalently, if the
secondary-noise floor is used as the comparison target, the maximum admissible
input clock ASD is larger by the same factor.
Figures~\ref{fig:simv2-PE-output}
and~\ref{fig:simv2-PE-strain} show the corresponding output-noise and
equivalent-strain spectra.

\begin{inlinefigure}
\centering
\includegraphics[width=\columnwidth,trim=45 15 75 40,clip]{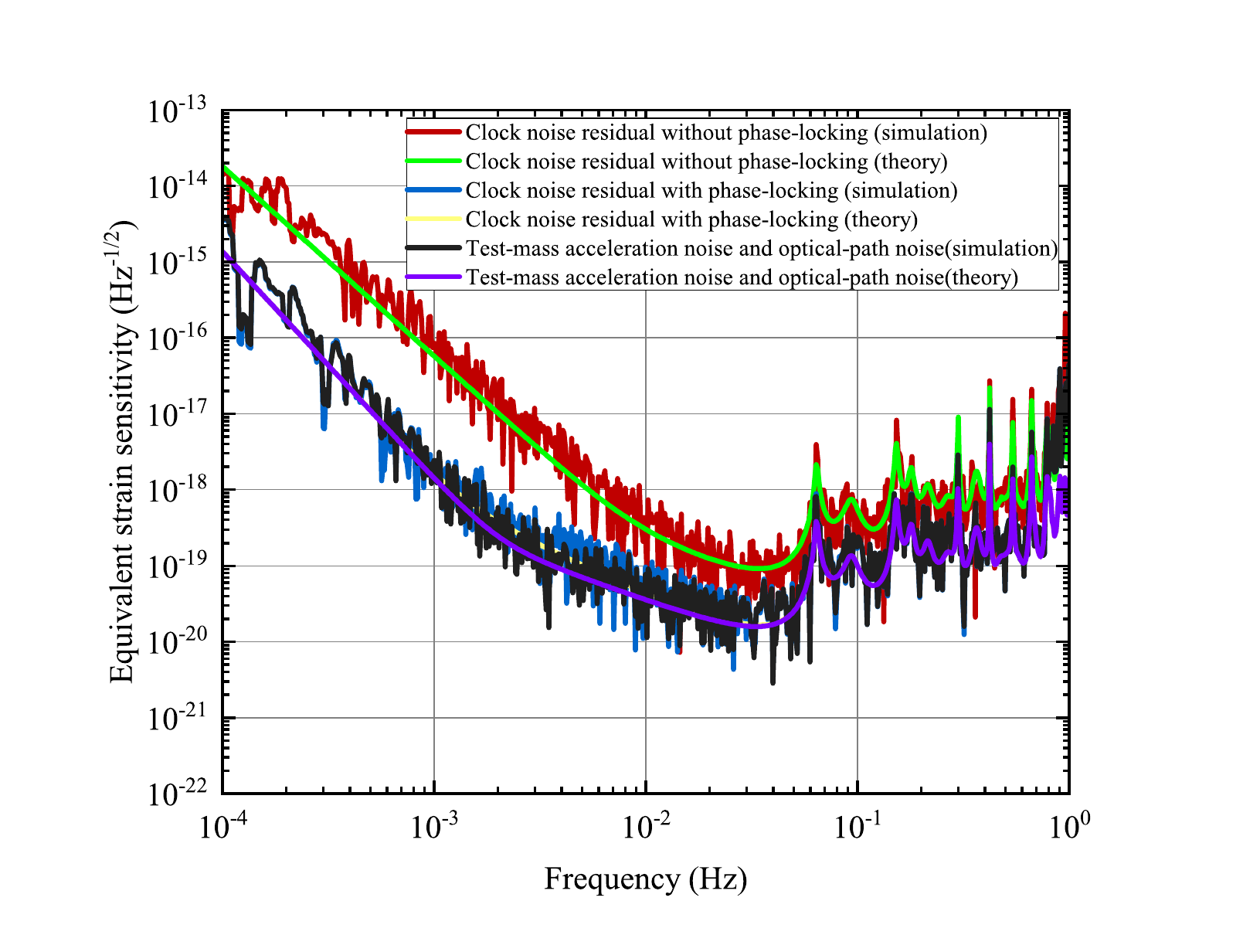}
\caption{\label{fig:simv2-PE-strain}
Equivalent-strain ASDs associated with $[PE]^{16}_{1}$, obtained by dividing
each output ASD by the absolute GW response of the corresponding realization.
The six curves have the same definitions as in
Fig.~\ref{fig:simv2-PE-output}.}
\end{inlinefigure}

For both simulated observables, $\mathcal G>1$ throughout the full
$10^{-4}$--$1~\mathrm{Hz}$ interval.
At a common transfer null, the ratio is understood by its continuous limit.
Within the adopted orbit average, frequency plan, and ideal phase-locking
model, the more conservative $[PE]^{16}_{1}$ result permits the input clock
ASD to be at least $5.351$ times larger than in the corresponding realization
without phase-locking.  These two examples numerically verify the lower
clock-noise residual obtained with phase-locking.  Together with the
catalog-wide transfer calculation, they support the reduction of clock-noise
coupling throughout the millihertz band.  The numerical factors
do not include finite phase-locking residuals, modulation-chain noise,
ranging errors, or uncertainties in the time-dependent couplings.

\section{Concluding remarks}\label{section7}

We have developed a clock-noise transfer and calibration framework directly
for a phase-locking configuration.  The carrier and sideband
measurements reduce the laser-noise sector to delayed copies of one master
process, while signed composite delay operators map any admissible two-branch
sequence to an ordered sideband template without commuting the time-dependent
delays.  This reorganization of the laser-noise sector does not attenuate the
master-laser fluctuation, whose cancellation still relies on TDI.  By contrast,
for a common input clock spectrum, each of the 45 distinct
second-generation geometric-TDI combinations with up to 16 links examined has
a smaller clock-noise residual after the TDI combination is formed and before
the final calibration template is applied in the phase-locking realization than
in the realization without phase-locking throughout
$0.1$--$10~\mathrm{mHz}$.  This result demonstrates a weaker coupling of
onboard-clock fluctuations to the observable in the phase-locking
configuration.  Numerical simulations of
$[X]^{16}_{1}$ and $[PE]^{16}_{1}$ reproduce the analytical spectra and
verify this reduction, while their calibrated residuals remain below the
secondary-noise floor in the ideal auxiliary-noise model.  These observables are representative examples rather
than limits of the construction, which applies whenever the propagation and
clock-coefficient closure conditions are satisfied.

\section*{Statements and Declarations}

\subsection*{Funding}

This work was supported by the National Key R$\&$D Program of China under Grant No. 2022YFC2204602 and by the National Natural Science Foundation of China under Grant No. 12405060.

\subsection*{Data and code availability}

The numerical data underlying the figures and the custom analysis code that
support the findings of this article are available from the corresponding
author upon reasonable request.

\subsection*{Competing interests}
The authors declare that they have no competing interests. They have no financial or non-financial interests relevant to the work reported in this article.

\renewcommand{\bibfont}{\normalfont\small}
\setlength{\bibsep}{3pt}
\bibliography{references_wang}

\end{multicols}
\end{document}